%% file: jpp-instructions.tex
\documentclass{jpp}
\makeatletter
\renewcommand{\fps@figure}{htbp}
\makeatother

\usepackage[utf8]{inputenc}
\usepackage[T1]{fontenc}

\usepackage{amsmath}
\usepackage{mathtools}
\usepackage{graphicx}
\usepackage{xcolor}
\usepackage{subcaption}
\usepackage{csquotes}

\usepackage{tikz}
\usetikzlibrary{matrix, arrows.meta}

\usepackage{microtype}
\usepackage[colorlinks=true, linkcolor=blue, citecolor=blue, pdfpagemode=UseOutlines, pdfauthor={Kaya E. Unalmis}, pdftitle={Fast, spectrally accurate, differentiable bounce-averaging}]{hyperref}
\usepackage{orcidlink}

\newcommand{\bookmarksection}[1]{%
  \section*{#1}%
  \phantomsection
  \addcontentsline{toc}{section}{#1}%
}

\providecommand{\aff}[1]{\textsuperscript{#1}}
\providecommand{\corresp}[1]{#1}
\providecommand{\email}[1]{\texttt{#1}}
\providecommand{\affiliation}[1]{#1}
\IfFileExists{latexml.sty}{\usepackage{latexml}}{}
\makeatletter
\@ifundefined{iflatexml}{\expandafter\newif\csname iflatexml\endcsname}{}
\makeatother

\input{kayamath}

\title{Spectrally accurate, reverse-mode differentiable bounce-averaging algorithm and its applications}
\author{Kaya Unalmis\aff{1} \corresp{\email{\href{mailto:kunalmis@alumni.princeton.edu}{kunalmis@alumni.princeton.edu}}}\orcidlink{0000-0002-1755-6764},
  Rahul Gaur\aff{2}\orcidlink{0000-0003-4367-9052},
  Rory Conlin\aff{3}\orcidlink{0000-0001-8366-2111},
  Dario Panici\aff{2}\orcidlink{0000-0003-0736-4360}
 \and Egemen Kolemen\aff{2,4,5} \corresp{\email{\href{mailto:ekolemen@princeton.edu}{ekolemen@princeton.edu}}}\orcidlink{0000-0003-4212-3247}}
\affiliation{\aff{1} Electrical and Computer Engineering, Princeton University, Princeton, NJ, USA
\aff{2} Mechanical and Aerospace Engineering, Princeton University, Princeton, NJ, USA
\aff{3} IREAP, University of Maryland, College Park, MD, USA
\aff{4} Andlinger Center for Energy and the Environment, Princeton University, Princeton, NJ, USA
\aff{5} Princeton Plasma Physics Laboratory, Princeton, NJ, USA}

\shortauthor{K. Unalmis, R. Gaur, R. Conlin, D. Panici and E. Kolemen}
\shorttitle{Fast, spectrally accurate, differentiable bounce-averaging}

\begin{document}
\maketitle
\begin{abstract}
	We present a fast, spectrally (exponentially) accurate, automatically differentiable bounce-averaging algorithm that is used to simplify kinetic models.
	Using this algorithm, implemented in the \texttt{DESC} stellarator optimisation suite, we can perform efficient optimisation of many objectives to improve stellarator performance, such as the effective ripple $\epsilon_{\mathrm{eff}}$ metric for the neoclassical transport coefficient in the low collisionality regime, energetic particle confinement, and turbulent transport.
	For the first time, we optimise a finite-beta stellarator to directly reduce neoclassical ripple transport using reverse-mode differentiation.
	This ensures the computational cost of differentiation is independent of the number of controllable parameters.
\end{abstract}

\input{1_introduction}
\input{2_neoclassical}
\input{3_method}
\input{4_optimization}
\input{5_summary}
\appendix
\input{A_open}
\input{B_ripple}
\newpage
\input{C_drift}
\input{D_pest_transform}

\input{published_references}
\end{document}

%% file: kayamath.tex
\iflatexml

	\DeclareRobustCommand{\e}{\mathrm{e}}
	\DeclareRobustCommand{\im}{\mathrm{i}}
	\DeclareRobustCommand{\cpi}{\mathrm{\pi}}
	
	\DeclareRobustCommand{\dl}{\mathrm{d}}
	\DeclareRobustCommand{\diff}[2]{\frac{\mathrm{d}{#1}}{\mathrm{d}{#2}}}
	\DeclareRobustCommand{\diffs}[2]{(\mathrm{d}{#1} / \mathrm{d}{#2})}
	
	\DeclareRobustCommand{\diffp}[2]{\frac{\partial {#1}}{\partial {#2}}}

	\makeatletter
	\DeclareRobustCommand{\abs}{\@ifstar{\abs@star}{\abs@nostar}}
	\newcommand{\abs@nostar}[1]{\lvert#1\rvert}
	\newcommand{\abs@star}[1]{\left\lvert#1\right\rvert}
	\DeclareRobustCommand{\norm}{\@ifstar{\norm@star}{\norm@nostar}}
	\newcommand{\norm@nostar}[1]{\lvert#1\rvert}
	\newcommand{\norm@star}[1]{\left\lvert#1\right\rvert}
	\DeclareRobustCommand{\ceil}{\@ifstar{\ceil@star}{\ceil@nostar}}
	\newcommand{\ceil@nostar}[1]{\lceil#1\rceil}
	\newcommand{\ceil@star}[1]{\left\lceil#1\right\rceil}
	\DeclareRobustCommand{\floor}{\@ifstar{\floor@star}{\floor@nostar}}
	\newcommand{\floor@nostar}[1]{\lfloor#1\rfloor}
	\newcommand{\floor@star}[1]{\left\lfloor#1\right\rfloor}
	\DeclareRobustCommand{\group}{\@ifstar{\group@star}{\group@nostar}}
	\newcommand{\group@nostar}[1]{\lparen#1\rparen}
	\newcommand{\group@star}[1]{\left\lparen#1\right\rparen}
	\DeclareRobustCommand{\groupbrack}{\@ifstar{\groupbrack@star}{\groupbrack@nostar}}
	\newcommand{\groupbrack@nostar}[1]{\lbrack#1\rbrack}
	\newcommand{\groupbrack@star}[1]{\left\lbrack#1\right\rbrack}
	
	\DeclareRobustCommand{\setSymbol}{%
		\mathchoice{\:}{\:}{\,}{\,}\vert
		\allowbreak
		\mathchoice{\:}{\:}{\,}{\,}
		\mathopen{}%
	}
	\DeclareRobustCommand{\set}{\@ifstar{\set@star}{\set@nostar}}
	\newcommand{\set@nostar}[1]{\lbrace#1\rbrace}
	\newcommand{\set@star}[1]{\left\lbrace#1\right\rbrace}
	\DeclareRobustCommand{\closeint}{\@ifstar{\closeint@star}{\closeint@nostar}}
	\newcommand{\closeint@nostar}[2]{\lbrack#1, #2\rbrack}
	\newcommand{\closeint@star}[2]{\left\lbrack#1, #2\right\rbrack}
	\DeclareRobustCommand{\openint}{\@ifstar{\openint@star}{\openint@nostar}}
	\newcommand{\openint@nostar}[2]{\lparen#1, #2\rparen}
	\newcommand{\openint@star}[2]{\left\lparen#1, #2\right\rparen}
	\DeclareRobustCommand{\clopenint}{\@ifstar{\clopenint@star}{\clopenint@nostar}}
	\newcommand{\clopenint@nostar}[2]{\lbrack#1, #2\rparen}
	\newcommand{\clopenint@star}[2]{\left\lbrack#1, #2\right\rparen}
	\DeclareRobustCommand{\openclint}{\@ifstar{\openclint@star}{\openclint@nostar}}
	\newcommand{\openclint@nostar}[2]{\lparen#1, #2\rbrack}
	\newcommand{\openclint@star}[2]{\left\lparen#1, #2\right\rbrack}
	\AtBeginDocument{\renewcommand{\vec}[1]{\boldsymbol{#1}}}
	\DeclareRobustCommand{\vecdot}{\vec{\cdot}}
	\AtBeginDocument{
		\let\oldnabla\nabla 
		\renewcommand{\nabla}{\boldsymbol{\oldnabla}}
	}
	\DeclareRobustCommand{\mean}{\@ifstar{\mean@star}{\mean@nostar}}
	\newcommand{\mean@nostar}[1]{\langle#1\rangle}
	\newcommand{\mean@star}[1]{\left\langle#1\right\rangle}
	\makeatother

\else

	\NewDocumentCommand{\e}{}{\mathrm{e}}
	\NewDocumentCommand{\im}{}{\mathrm{i}}
	\NewDocumentCommand{\cpi}{}{\mathrm{\pi}}
	
	\NewDocumentCommand{\dl}{}{\mathrm{d}}
	\NewDocumentCommand{\diff}{mm}{\frac{\mathrm{d}{#1}}{\mathrm{d}{#2}}}
	\NewDocumentCommand{\diffs}{mm}{(\mathrm{d}{#1} / \mathrm{d}{#2})}
	
	\NewDocumentCommand{\diffp}{mm}{\frac{\partial {#1}}{\partial {#2}}}

	\DeclarePairedDelimiter{\abs}{\lvert}{\rvert}
	\DeclarePairedDelimiter{\norm}{\lvert}{\rvert}
	\DeclarePairedDelimiter{\ceil}{\lceil}{\rceil}
	\DeclarePairedDelimiter{\floor}{\lfloor}{\rfloor}
	\DeclarePairedDelimiter{\group}{\lparen}{\rparen}
	\DeclarePairedDelimiter{\groupbrack}{\lbrack}{\rbrack}
	
	\NewDocumentCommand{\setSymbol}{o}{
		\mathchoice{\:}{\:}{\,}{\,}\IfValueT{#1}{#1}\vert
		\allowbreak
		\mathchoice{\:}{\:}{\,}{\,}
		\mathopen{}}
	\DeclarePairedDelimiterX{\set}[1]{\lbrace}{\rbrace}{%
		
		#1
	}
	\DeclarePairedDelimiterX{\closeint}[2]{\lbrack}{\rbrack}{#1, #2}
	\DeclarePairedDelimiterX{\openint}[2]{\lparen}{\rparen}{#1, #2}
	\DeclarePairedDelimiterX{\clopenint}[2]{\lbrack}{\rparen}{#1, #2}
	\DeclarePairedDelimiterX{\openclint}[2]{\lparen}{\rbrack}{#1, #2}
	\AtBeginDocument{\RenewDocumentCommand{\vec}{m}{\boldsymbol{#1}}}
	\NewDocumentCommand{\vecdot}{}{\vec{\cdot}}
	\AtBeginDocument{
		\let\oldnabla\nabla 
		\RenewDocumentCommand{\nabla}{}{\boldsymbol{\oldnabla}}
	}
	\DeclarePairedDelimiter{\mean}{\langle}{\rangle}

\fi

%% file: 1_introduction.tex
\section{Introduction}
Stellarators, first conceived by~\cite{spitzer1958stellarator}, represent a distinct approach to magnetic confinement fusion that offers unique advantages over tokamaks. These toroidal devices achieve plasma confinement through external magnetic fields rather than through plasma current, providing greater design flexibility and operational stability.
The absence of a continuous toroidal symmetry allows for magnetic field optimisation through boundary shaping, which helps minimise the net toroidal current and thereby avoids current-driven plasma disruptions that plague tokamak operation~\citep{helander2014theory}.

The design of optimal stellarator configurations is a complex optimisation problem involving hundreds of degrees of freedom.
Traditional optimisation approaches have evolved significantly over the past few decades. \texttt{VMEC} (Variational Moments Equilibrium Code), developed by~\cite{hirshman1983steepest_VMEC}, served as the foundation for stellarator optimisation.
Building upon~\texttt{VMEC}, several frameworks have emerged: \texttt{STELLOPT}~\citep{spong1998stellopt, stellopt},
which implements a suite of physics-based optimisation criteria, \texttt{ROSE}~\citep{drevlak2018optimisation}, which focuses on coil optimisation and engineering constraints, and more recently, \texttt{SIMSOPT}~\citep{landreman2021simsopt}.
In \texttt{DESC}~\citep{dudt2020desc, conlin2023desc, Dudt_Conlin_Panici_Kolemen_2023, panici2023desc, desc_git}, unlike previous optimisers, it is not necessary to re-solve the magnetohydrodynamic (MHD) force balance equation at each optimisation step \citep{Conlin_Kim_Dudt_Panici_Kolemen_2024}.
Additional objectives that depend on equilibrium force balance can be optimised simultaneously while ensuring force balance.

Traditional approaches to stellarator optimisation rely on finite difference techniques.
Such techniques are low-order accurate and hinder the ability of the optimiser to find good solutions.
Furthermore, finite difference techniques require computing the objective function multiple times to estimate the derivative in the direction of each optimisable parameter; this is infeasible when the number of parameters is large.
In contrast, adjoint methods and automatic differentiation \citep{sapienza2025differentiableprogrammingdifferentialequations} can compute derivatives with respect to all optimisable parameters with a computational cost that is comparable to the cost of a single objective function evaluation.
These methods have greatly improved the ability to solve inverse design problems for stellarators (\citealp*{Antonsen_Paul_Landreman_2019}; \citealp{Paul_Abel_Landreman_Dorland_2019}; \citealp*{Paul_Landreman_Antonsen_2021}; \citealp{desc_git}).

We present an automatically differentiable bounce-averaging algorithm that is used to simplify kinetic models such as drift and gyrokinetics that study phenomena at time scales longer than the bounce orbit time.
This algorithm has been implemented in the \texttt{DESC} optimisation suite.
Previous works \citep{MATSUDA1986197, nemov1999evaluation, 10.1063/1.2912456, Kernbichler_2016, Petrov_2016, stellopt, VELASCO2020109512, Velasco_2021} have used bounce-averaging to accelerate the solution of Fokker--Planck equations.
However, such works are incompatible with automatic differentiation.
Moreover, their computation is discretised with lower-order accuracy than in this work.
This work enables fast optimisation to improve stellarator performance with exponential accuracy.

In \S~\ref{sec:neoclassical}, we present an application of bounce-averaging to compute the neoclassical transport coefficient in the low collisionality regime where the transport coefficients increase with decreasing collision frequency.
Then, in \S~\ref{sec:algo}, we describe our numerical methods to compute bounce-averaged objectives for optimisation.
In \S~\ref{sec:optimization}, we apply this framework to optimise against neoclassical transport.
In \S~\ref{sec:conclusion}, we conclude this work and explain how it can be extended.

%% file: 2_neoclassical.tex
\section{Neoclassical model of plasma}
\label{sec:neoclassical}
Our study concerns configurations where magnetic field lines lie on closed, nested toroidal surfaces, known as flux surfaces.
We label these surfaces with their enclosed toroidal flux $\psi$.
Such a divergence-free magnetic field may be written in the Clebsch form~\citep{d2012flux}, showing that curves of constant \((\psi, \alpha)\) trace field lines,
\begin{equation}
	\vec{B} = \nabla \psi \times \nabla \alpha.
	\label{eqn:Div-free-B1}
\end{equation}

The dynamics of a magnetised hot plasma differ significantly from those of an unmagnetised fluid.
Unlike isotropic hard-sphere collisions that govern the behaviour of an uncharged fluid, a plasma behaves differently in directions perpendicular and parallel to the magnetic field lines because of Coulomb collisions.
In magnetised plasmas, particles traverse helical trajectories, gyrating around magnetic field lines and drifting across them.
The classical transport model assumes a simplistic view of particle collisions and does not adequately incorporate the effects of these drifts.
To properly account for these drifts, trapped and passing particles, and the magnetic geometry, we use the neoclassical transport theory.

There are three fundamental length and time scales relevant to magnetised plasmas.
The time scales correspond to the particle transit frequency $v_{\mathrm{th,s}}/L_{B}$, where $v_{\mathrm{th,s}} = (2 T_{s}/m_s)^{1/2}$ is the thermal speed, the Coulomb collision frequency $\nu_{ss^{\prime}} \propto T^{-3/2}$ and the gyration frequency $\Omega_{s} = Z_s e \norm{B}/(m_s c)$, where $s, s^{\prime}$ are the species of interest, \(Z_s e\) is the charge, \(m_s\) is the mass and \(c\) is the speed of light.
For each time scale, the corresponding length scales are the gradient scale length of the magnetic field $L_{B}$, the mean free path $\lambda_{\mathrm{mfp}}$ and the gyroradius $r_{\text{gyro},s} = v_{\mathrm{th}, s}/\Omega_s$, respectively.
In a magnetised plasma,
\begin{align}
	\nu_{ss^{\prime}} \sim \frac{v_{\mathrm{th, s}}}{L_{B}} & \ll \Omega_s, \label{eq:gryospeed} \\
	\lambda_{\mathrm{mfp}} \sim L_{B}                       & \gg r_{\text{gyro},s}.
\end{align}

Using a random walk estimate, we can calculate the classical heat transport coefficient in the perpendicular direction as $D_{\perp} \sim \nu_{ss^{\prime}} r_{\text{gyro},s}^2 \sim T^{-1/2}$~\citep{helander2005collisional}, whereas, using neoclassical theory, we have $\Delta r \sim r_{\text{gyro},s} \norm{B}/\norm{B}_{\text{poloidal}}$ with $\norm{B}$ and $\norm{B}_{\text{poloidal}}$ given by the total and poloidal magnetic field strength, respectively.
The transport coefficient is then $D_{\perp} \sim \nu_{ss^{\prime}} r_{\text{gyro},s}^2 \norm{B}^2/\norm{B}_{\text{poloidal}}^2 \sim T^{-1/2} \norm{B}^2/\norm{B}_{\text{poloidal}}^2$.
Note that the ratio $\norm{B}/\norm{B}_{\text{poloidal}}$ strongly depends on the magnetic field geometry and significantly affects the regime of neoclassical transport.

Magnetised plasmas can be weakly or strongly collisional
as defined by the collisionality $\nu_{\star} = L_{B}/\lambda_{\mathrm{mfp}}$.
In a strongly collisional plasma, particles undergo frequent collisions without covering significant distances along a magnetic field line, \textit{i.e.} $\nu_{\star} \gg 1$.
Conversely, in a weakly collisional plasma, particles can traverse a significant distance before colliding, \textit{i.e.} $\nu_{\star} \ll 1$.
Stellarator plasmas in practical applications tend to be weakly collisional.

Based on the stellarator geometry, the weak collisionality regime can be further partitioned into the banana or plateau regime according to the reciprocal of the aspect ratio $\epsilon \sim \iota^{-1} \norm{B}_{\text{poloidal}} / \norm{B}$, where \(\iota\) is the rotational transform \citep{helander2014theory}.
Most stellarators lie in the regime where
$\nu_{\star} \ll \epsilon^{3/2}$.
This categorisation is illustrated in figure~\ref{fig:neoclassical-illustration}.

The standard neoclassical theory first enabled computation of the neoclassical transport coefficients in the low collisionality regime for a simplified model of the magnetic field.
This analysis was later extended to stellarator magnetic fields \citep{Kovrizhnykh_1984, Ochs_2025}.
The following section outlines this process in one regime of interest to stellarator equilibrium optimisation.

\vspace{-0.35cm}
\begin{figure}
	\centering
	\begin{tikzpicture}
		\tikzset{
		block/.style={
				rectangle,
				rounded corners=3mm,
				fill=blue!4,
				text width=8.5em,
				align=center,
			},
		arrow/.style={
		draw,
		thick,
		color=black!50,
		arrows={-Stealth[length=2mm, width=2mm]}
		}
		}
		\matrix (m) [
		matrix of nodes,
		row sep=0.3cm,
		column sep=0.4cm,
		nodes={block}
		]
		{
		& |[name=ionized]| {Fully ionised plasma} & \\
		|[name=mag]| {Magnetised \\ $r_{\text{gyro},s}/L_{B} \ll 1$} & & |[name=unmag]| {Un-magnetised \\ $r_{\text{gyro},s}/L_{B} \sim 1$} \\
		|[name=ps]| {Collisional \\ $\lambda_{\text{mfp}}/L_{B} \ll 1$ \\ \textbf{Pfirsch--Schl{\"u}ter} \\ (minimal trapping)} & |[name=trapped]| {Weakly collisional \\ $\lambda_{\text{mfp}}/L_{B} \gg 1$ \\ \textbf{Trapped orbits}} & \\
		& |[name=banana]| {Weakly affected \\ \textbf{Banana}} & |[name=plateau]| {Strongly affected \\ \textbf{Plateau}} \\
		};

		\path[arrow] (ionized) -- (mag);
		\path[arrow] (ionized) -- (unmag);
		\path[arrow] (mag) -- (ps);
		\path[arrow] (mag) -- (trapped);
		\path[arrow] (trapped) -- (banana);
		\path[arrow] (trapped) -- (plateau);
	\end{tikzpicture}
	\caption{A schematic categorising neoclassical transport.}
	\label{fig:neoclassical-illustration}
\end{figure}

\vspace{-0.37cm}
\subsection{Effective ripple}
In the low collision limit $\nu_{\star} \ll \epsilon^{3/2}$, the neoclassical model studies the plasma distribution \(f\) determined by a simplified Boltzmann equation known as the drift-kinetic equation.
For a particle with mass \(m\), let \(\vec{v}_{\parallel}\) and \(\vec{v}_{\perp}\) be the velocity parallel and orthogonal, respectively, to the unit vector magnetic field \(\vec{b}\).
In the drift-kinetic equation, the velocity space may be parametrised with three independent coordinates: the total energy \(E\), the magnetic moment \(\mu = m \norm{v_{\perp}}^{2}/(2 \norm{B})\) and the gyrophase angle.
In this treatment, the equation is averaged over the gyrophase angle.
We seek a steady-state solution and linearise the distribution of guiding centres \(f = f_0 + f_1\) into a background \(f_0\) that is Maxwellian in velocity and a higher order correction \(f_1\).
Thus, the background is parametrised in velocity space with \(E\) and the higher order correction with \((E, \mu)\).
The linearised drift-kinetic equation reduces to the following partial differential equation (PDE) \citep{Abel_2013}:
\begin{gather}
	\mathcal{C}[f] = \vec{v}_{\mathrm{D}s} \vecdot \nabla f_0 + \norm{v_{\parallel}} \vec{b} \vecdot \nabla f_1, \label{eqn:Neoclassical-equation} \\
	\vec{v}_{\mathrm{D}s} = \frac{\norm{v_{\parallel}}^2}{\Omega_s} \vec{b} \times (\vec{b} \vecdot \nabla\vec{b}) + \frac{\norm{v_{\perp}}^2}{2 \Omega_s} \frac{\vec{b} \times \nabla \norm{B}}{\norm{B}} + \vec{v}_{\text{Baños}}. \label{eqn:driftv}
\end{gather}
The electric field is neglected in this section as we focus on the \(1/\nu\) collisionality regime.

\begin{figure}
	\centering
	\includegraphics[width=\textwidth]{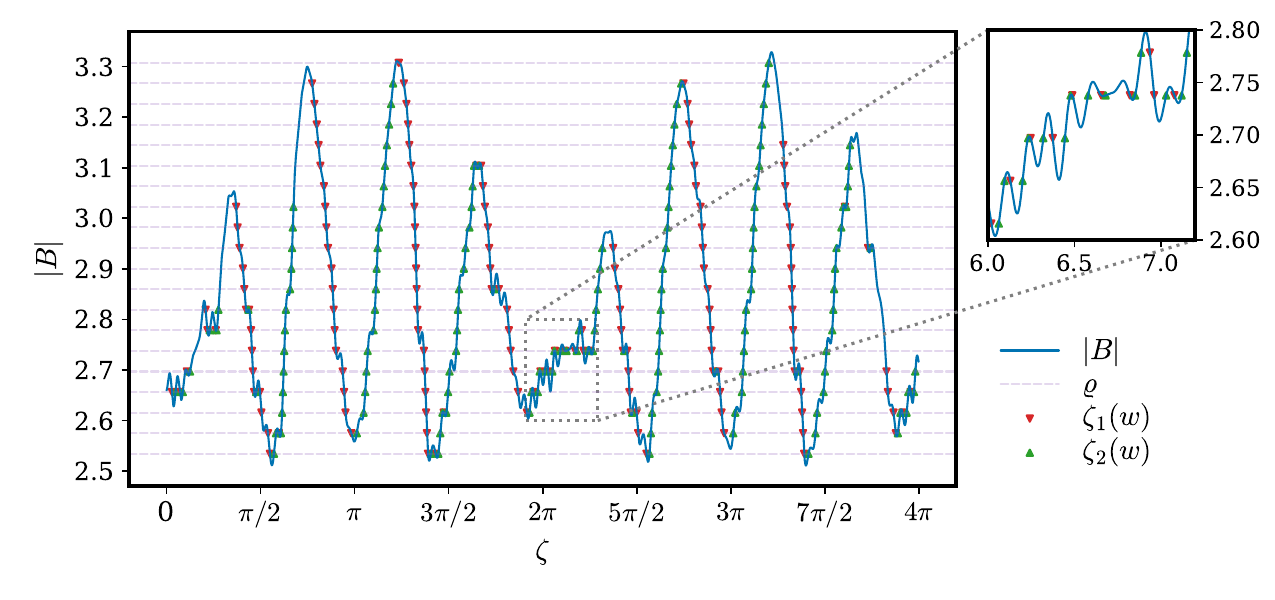}
	\caption{Bounce points within \((\zeta_1, \zeta_2) = (0, 4 \cpi)\) on the field line \((\psi, \alpha) = (\psi_{\text{plasma boundary}}, 0)\) for a mesh of \(\varrho\) values on a W7-X stellarator. For a given \(\varrho\) marked by a horizontal line, \(\norm{v_{\parallel}} = 0\) at the bounce points marked by triangles. The plasma distribution vanishes in the hypograph of \(\norm{B}\).}
	\label{fig:bouncepointvisual}
\end{figure}

To reduce neoclassical transport, one may minimise the radial particle flux density,
\begin{equation}
	\Gamma = \int \dl{^3\vec{v}} \; f_1 \vec{v}_{\mathrm{D}s}\vecdot \nabla \psi. \label{eq:orrad}
\end{equation}
Appendix \ref{sec:neoderivation} shows a derivation to extract a dimensionless quantity \(\Gamma_0\) \eqref{eq:epsnoconst} for the optimisation objective which is proportional to the flux surface average \(\mean{\Gamma}\) \eqref{eq:surfcompact}.
\begin{align}
	\mean{\Gamma}                 & = -\Gamma_0 \frac{2^{3/2} \cpi c^2}{3^2 e^2 m^{3/2}} \int_{0}^{\infty} \dl E \; \frac{E^{5/2}}{\nu} \frac{\partial f_0}{\partial \psi} \label{eq:biggamma}                                                                                                                                        \\
	\Gamma_0                      & = \group*{\int_{0}^{2\cpi} \dl \alpha \int_{\norm{B}_{\text{min}}}^{\norm{B}_{\text{max}}} \frac{\dl \varrho}{\varrho^3} \; \sum_{w} \frac{I_1^2}{I_2} }  \group*{\int_{0}^{2\cpi} \dl \alpha \int_{\zeta_1}^{\zeta_2} \frac{\dl \zeta}{\vec{B} \vecdot \nabla \zeta}}^{-1} \label{eq:epsnoconst} \\
	I_1(\psi, \alpha, \varrho, w) & = \int_{\zeta_1(w)}^{\zeta_2(w)} \frac{\dl \zeta}{\vec{B}\vecdot \nabla \zeta} \group{1 -  \norm{B} / \varrho}^{1/2} \group{4 \varrho / \norm{B} - 1} \norm{\nabla\psi} \kappa_{\mathrm{G}}                                                                                                       \\
	I_2(\psi, \alpha, \varrho, w) & = \int_{\zeta_1(w)}^{\zeta_2(w)} \frac{\dl \zeta}{\vec{B}\vecdot \nabla \zeta} \group{1 - \norm{B} / \varrho}^{1/2}
\end{align}
The quantity \(\kappa_{\mathrm{G}}\) is the geodesic curvature of the field line \eqref{eq:kappag} and the velocity space coordinate \(\varrho\) is defined as
\begin{equation}
	\varrho = E / \mu.
\end{equation}
The index \(w\) labels the well with boundaries \(\zeta_1(w)\) and \(\zeta_2(w)\) where a bouncing particle is trapped.
These boundaries are referred to as bounce points.
Only the particles which are trapped within the interval \(\closeint{\zeta_1}{\zeta_2}\) are considered so that \(\zeta_1 \leq \min_w \zeta_1(w) \) and \(\max_w \zeta_2(w) \leq \zeta_2\).
An illustration is shown in figure \ref{fig:bouncepointvisual}.

In axisymmetric configurations, integration along the field line for a single poloidal transit between two global maxima of \(\norm{B}\) is sufficient for convergence of \(\Gamma_0\).
On irrational magnetic surfaces, it is sufficient to integrate along a single field line \cite[\S~4.9]{d2012flux}.
On a rational or near-rational surface in non-axisymmetric configurations, it is necessary to integrate along multiple field lines until the surface is covered sufficiently.

The effective ripple modulation amplitude \(\epsilon_{\text{eff}}\) is related to \(\Gamma_0\) as follows:
\begin{equation}
	\epsilon_{\text{eff}}^{3/2} = \frac{\cpi}{2^{7/2}}\frac{(B_0 R_0)^2}{\mean{\norm{\nabla \psi}}^{2}}  \Gamma_0,
\end{equation}
where $B_0$ is a background magnetic field typically chosen to be \(\norm{B}_{\text{max}}\) and $R_0$ is the average major radius of the stellarator.
A reason $\epsilon_{\text{eff}}$ is preferred to \(\Gamma_0\) as an optimisation objective is that the latter vanishes near the magnetic axis, which reduces the ability to distinguish between good and bad configurations.
Since $\epsilon_{\text{eff}}$ depends only on geometry, reducing it by varying the plasma boundary can reduce the radial neoclassical loss of trapped particles.

%% file: 3_method.tex
\section{Algorithm} \label{sec:algo}
We briefly describe a few fundamental parts of our algorithm.
Section \ref{sec:bounceintintro} discusses the bounce integral in more detail.
In \S~\ref{sec:specquad}, we describe efficient quadrature used for these integrals.
Section \ref{sec:invertalgo} discusses our inverse method to solve the ideal MHD equation.
In \S\S~\ref{sec:mitoff}--\ref{sec:jacmap}, we describe our approach to obtain data along field lines.

To motivate the need for an efficient algorithm, let us estimate the computational cost of bounce-averaging with a blunt approach to the computation.
After discretising to $N_s$ field lines, where each field line is followed over $N_w$ magnetic wells for each of $N_{\varrho}$ pitch angles, there will be $\mathcal{O}(N_s N_w N_{\varrho})$ bounce integrals.
With $N_{q}$ quadrature points each, the integrand is evaluated at $\mathcal{O}(N_s N_w N_{\varrho} N_q) \sim 10^{8}$ points.
Furthermore, the path of integration is unknown \textit{a priori} because the field lines move during optimisation.
Finding the position of the field lines on a known grid may involve \(N_i\) Newton iterations for each point.
With \(N_c\) spectral coefficients used to approximate the map on which that root-finding is done, the cost grows to \(\mathcal{O}(N_c N_i N_s N_w N_{\varrho} N_q)\).
Moreover, the memory required to reverse-mode differentiate the objective grows linearly with the problem size.

\subsection{Bounce integral} \label{sec:bounceintintro}
The bounce integral of \(x\) may be written as a time-weighted integral over the trajectory of the particle along its bounce orbit \citep[\S~2]{10.1063/5.0160282}.
Since the dynamics parallel to the field lines dominate, the particle trajectory may be approximated to follow field lines by parametrising time \(t\) as the distance along a field-line following coordinate.
Since the magnetic moment is an adiabatic invariant for which the gyro-average of \(\dl \mu / \dl t\) is approximately zero, the pitch angle of a bouncing particle stays nearly constant over the time scale to complete bounce orbits when energy is conserved.
Labelling the boundaries \(\zeta_1(w)\) and \(\zeta_2(w)\) of magnetic well \(w\) where the parallel velocity vanishes,
using the streamline property in curvilinear coordinates,
\begin{equation}
	\norm{v_{\parallel}} \dl t = \frac{\dl \zeta}{\vec{b} \vecdot \nabla \zeta},
\end{equation}
and \(\norm{v_{\parallel}}^2 = (2 E / m) (1 - \norm{B} / \varrho)\), the integral may be written as
\begin{equation}
	\overline{x}(\psi, \alpha, \varrho, w) = \frac{m^{1/2}}{(2 E)^{1/2}} \int_{\zeta_{1}(w)}^{\zeta_{2}(w)} \frac{\dl \zeta}{\vec{b}\vecdot \nabla \zeta} (1 - \norm{B} / \varrho )^{-1/2} x. \label{eq:vdrift}
\end{equation}

More generally, integrals between bounce points involve a map \(g\), smooth in \(\zeta\), weighted by a map with behaviour matching \(\norm{v_{\parallel}}^{\eta}\) near the bounce points,
\begin{equation}
	\int_{\zeta_{1}(w)}^{\zeta_{2}(w)} \dl \zeta \, \norm{v_{\parallel}}^{\eta} g(\psi, \alpha, \zeta, \varrho, E), \qquad  \eta \in \set{-1, 1}. \label{eq:badint}
\end{equation}

\subsection{Quadrature} \label{sec:specquad}
Gaussian quadrature approximates \(\int_{-1}^{1} \dl \zeta \, \varsigma g(\zeta) \approx \sum_{i=1}^{N_q} \sigma_i g(\zeta_i)\) for some weight \(\varsigma\) positive and continuous in the interior by approximating \(g\) with its Hermite interpolation polynomial and choosing \(\sigma_i, \zeta_i\) to avoid evaluating the derivative \citep{Suli}.
For integrable \eqref{eq:badint}, we can construct such a quadrature for \(\varsigma\) matching the non-polynomial behaviour of \(\norm{v_{\parallel}}^{\eta}\) or, more generally, employ a change of variable whose Jacobian vanishes slowly near singularities such that the integrand can then be approximated by a low-degree polynomial.
In the latter approach, the transformation should also be mild enough to prevent unnecessary clustering of quadrature points that would increase the condition number of the problem.

\begin{figure}
	\bigskip
	\centering
	\begin{subfigure}[t]{0.4325\textwidth}
		\centering
		\includegraphics[height=3.5cm]{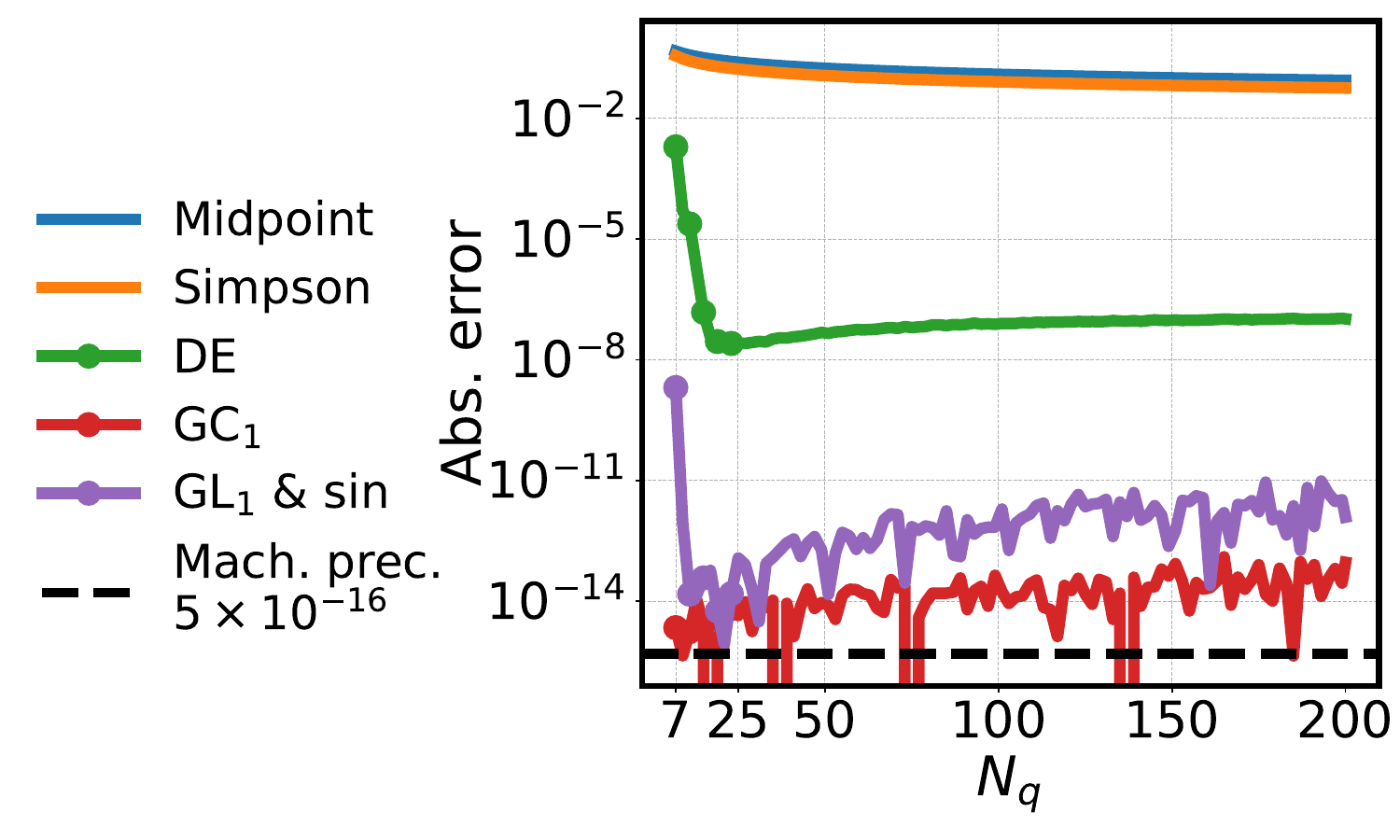}
		\caption{$k = 0.25$}
	\end{subfigure}
	\hfill
	\begin{subfigure}[t]{0.2925\textwidth}
		\centering
		\includegraphics[height=3.5cm]{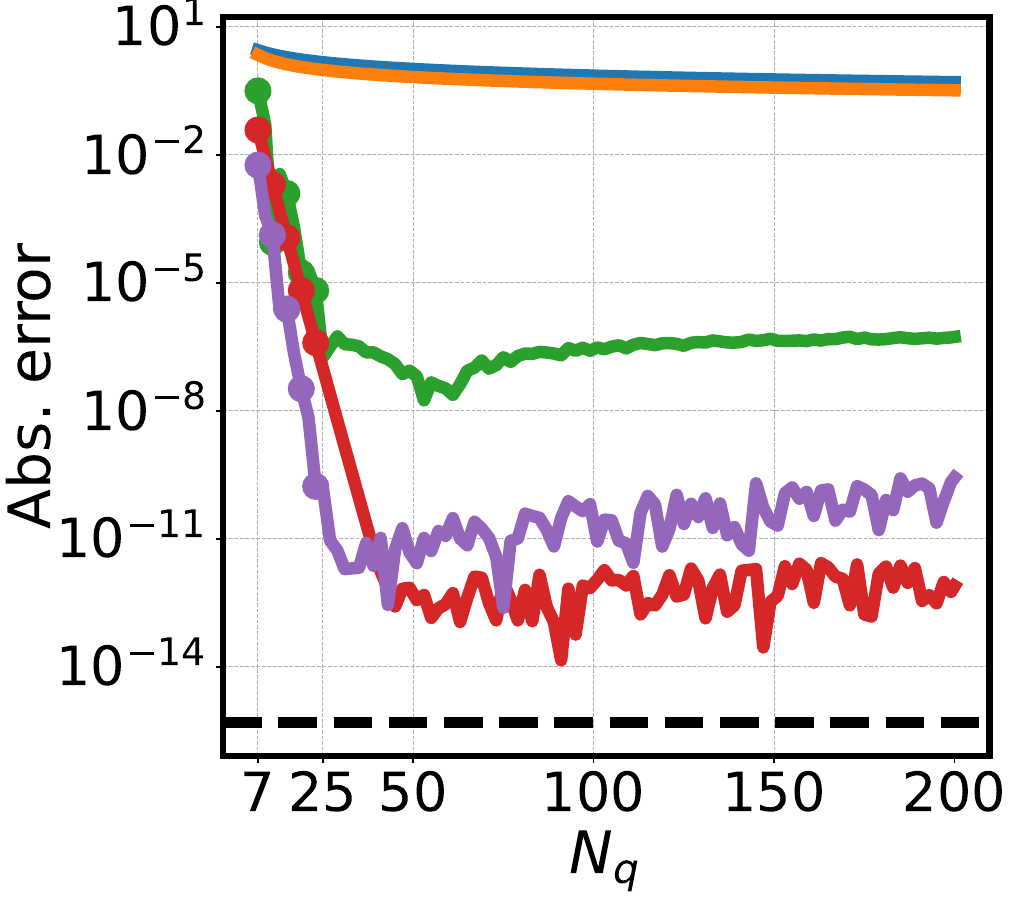}
		\caption{$k=0.999$}
	\end{subfigure}
	\hfill
	\begin{subfigure}[t]{0.26\textwidth}
		\centering
		\includegraphics[height=3.5cm]{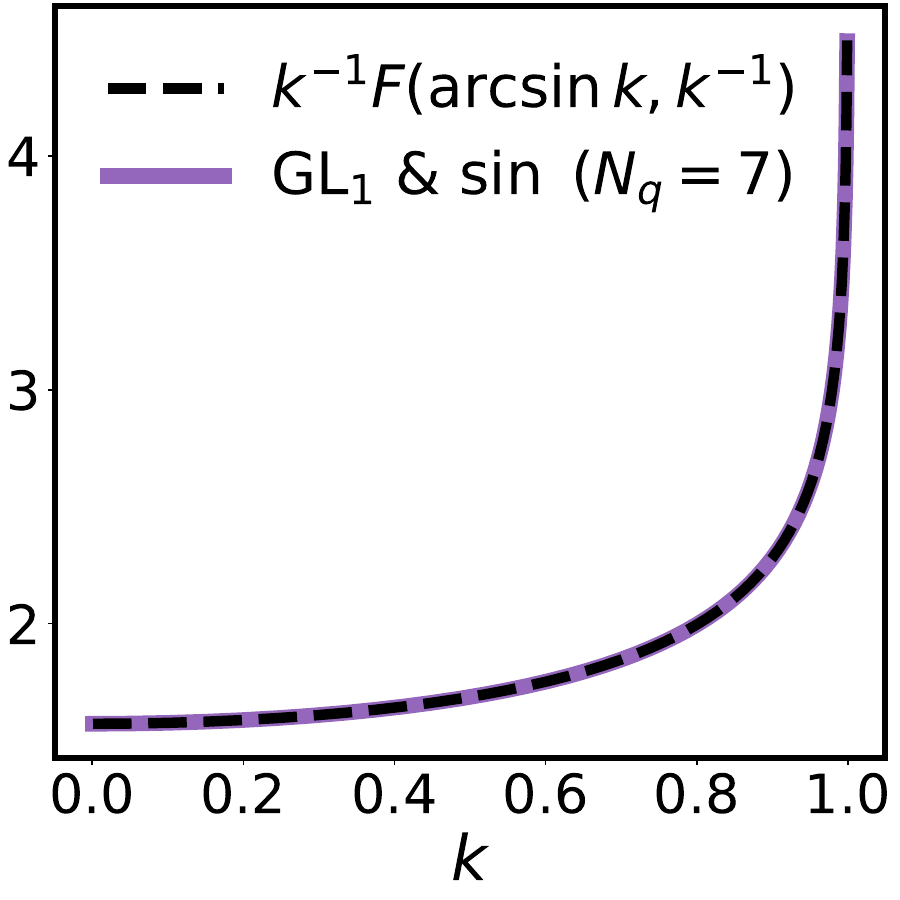}
		\caption{Elliptic F}
	\end{subfigure}
	\caption{Convergence of quadratures for \protect{\eqref{eqn:Elliptic-integral-F}}. $\mathrm{GC}_1$ and $\mathrm{GL}_1$ show spectral convergence whereas midpoint, Simpson and double exponential quadratures hit floating point plateaus early.}
	\label{fig:app-F-quad-compare}
	\bigskip

	\begin{subfigure}[t]{0.4325\textwidth}
		\centering
		\includegraphics[height=3.5cm]{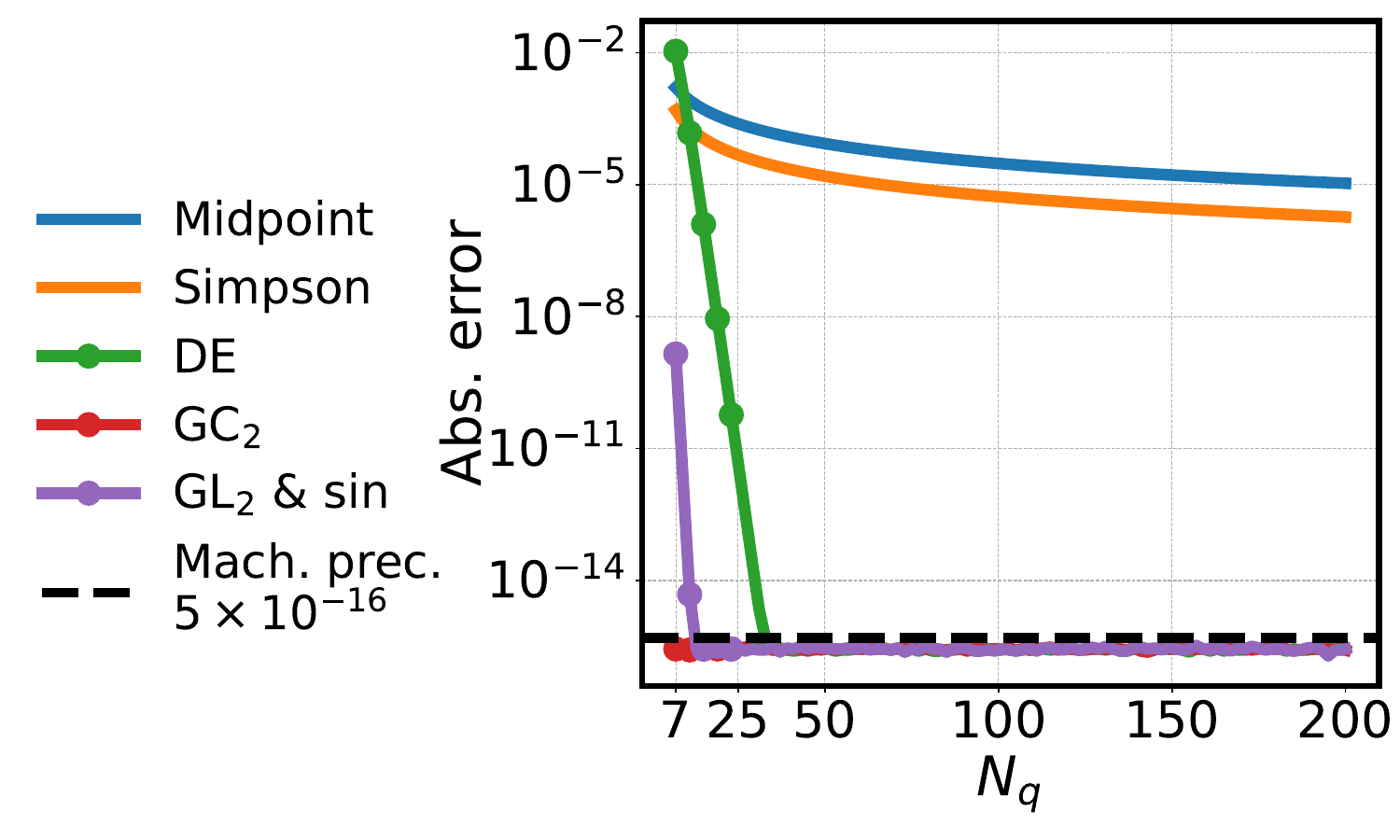}
		\caption{$k = 0.25$}
	\end{subfigure}
	\hfill
	\begin{subfigure}[t]{0.2925\textwidth}
		\centering
		\includegraphics[height=3.5cm]{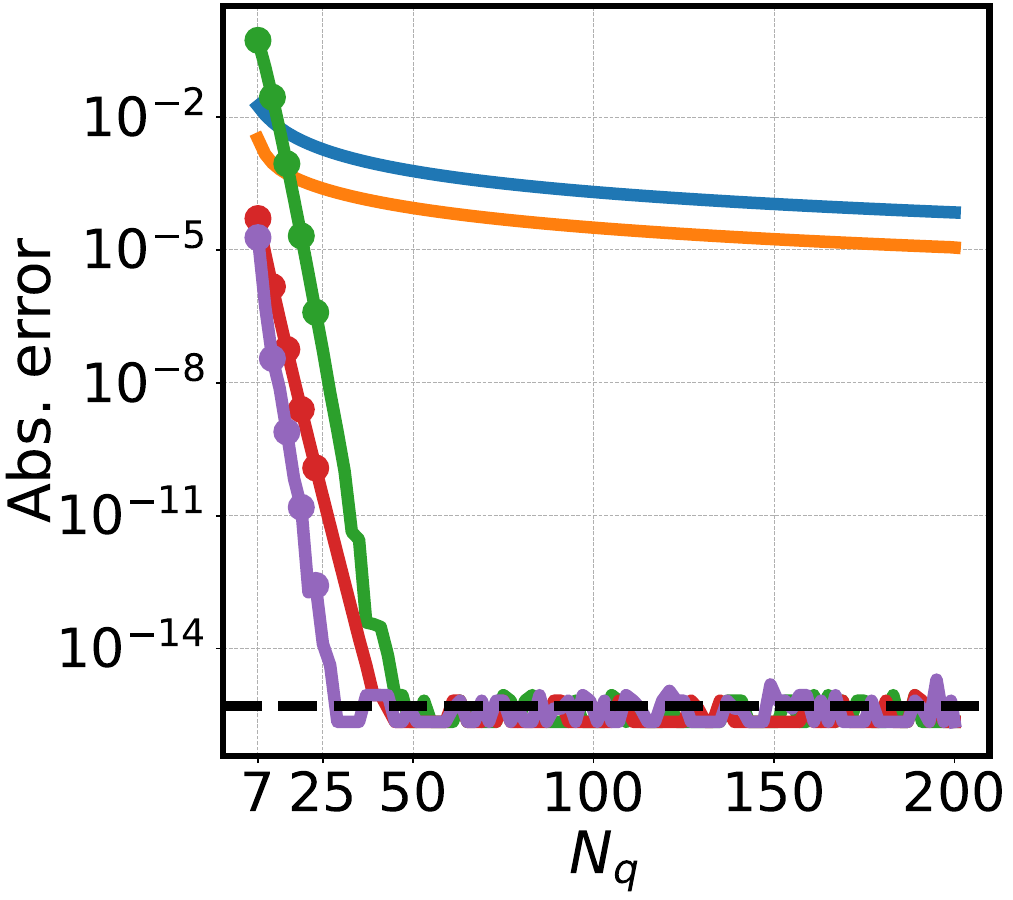}
		\caption{$k=0.999$}
	\end{subfigure}
	\hfill
	\begin{subfigure}[t]{0.26\textwidth}
		\centering
		\includegraphics[height=3.5cm]{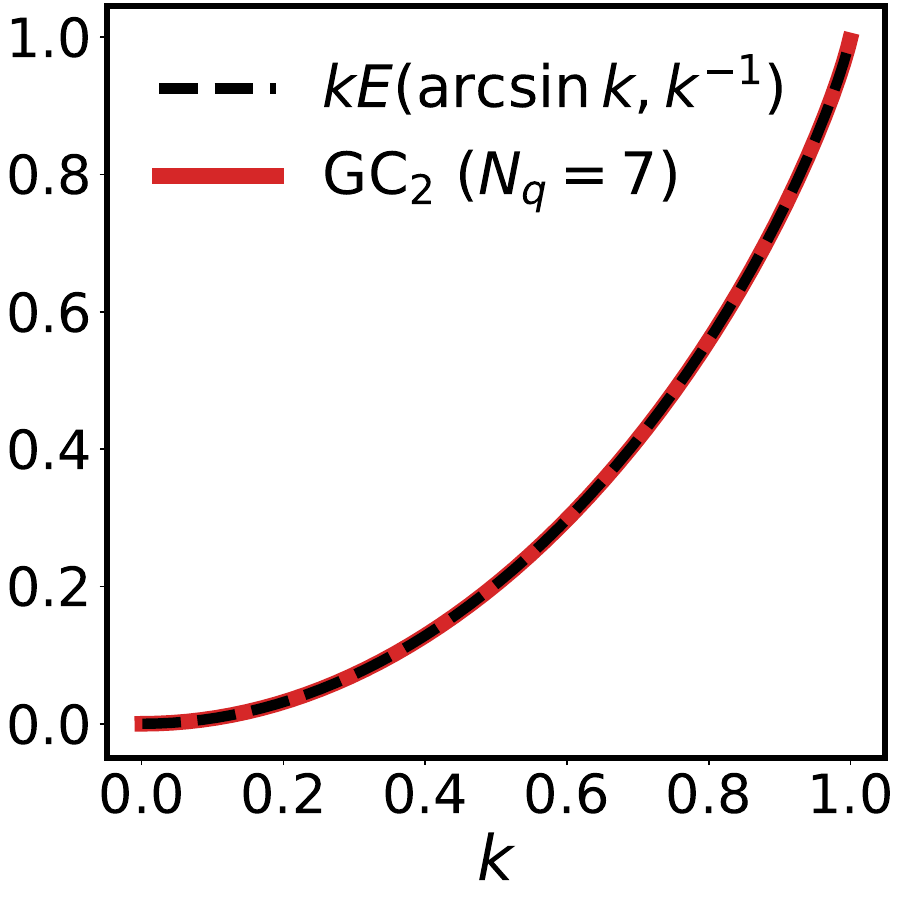}
		\caption{Elliptic E}
	\end{subfigure}
	\caption{Convergence of quadratures for \protect{\eqref{eqn:Elliptic-integral-E}}. $\mathrm{GC}_2$ and $\mathrm{GL}_2$ show spectral convergence.}
	\label{fig:app-E-quad-compare}
\end{figure}

\begin{figure}
	\bigskip
	\centering
	\begin{subfigure}[t]{0.48\textwidth}
		\centering
		\includegraphics[height=3.85cm]{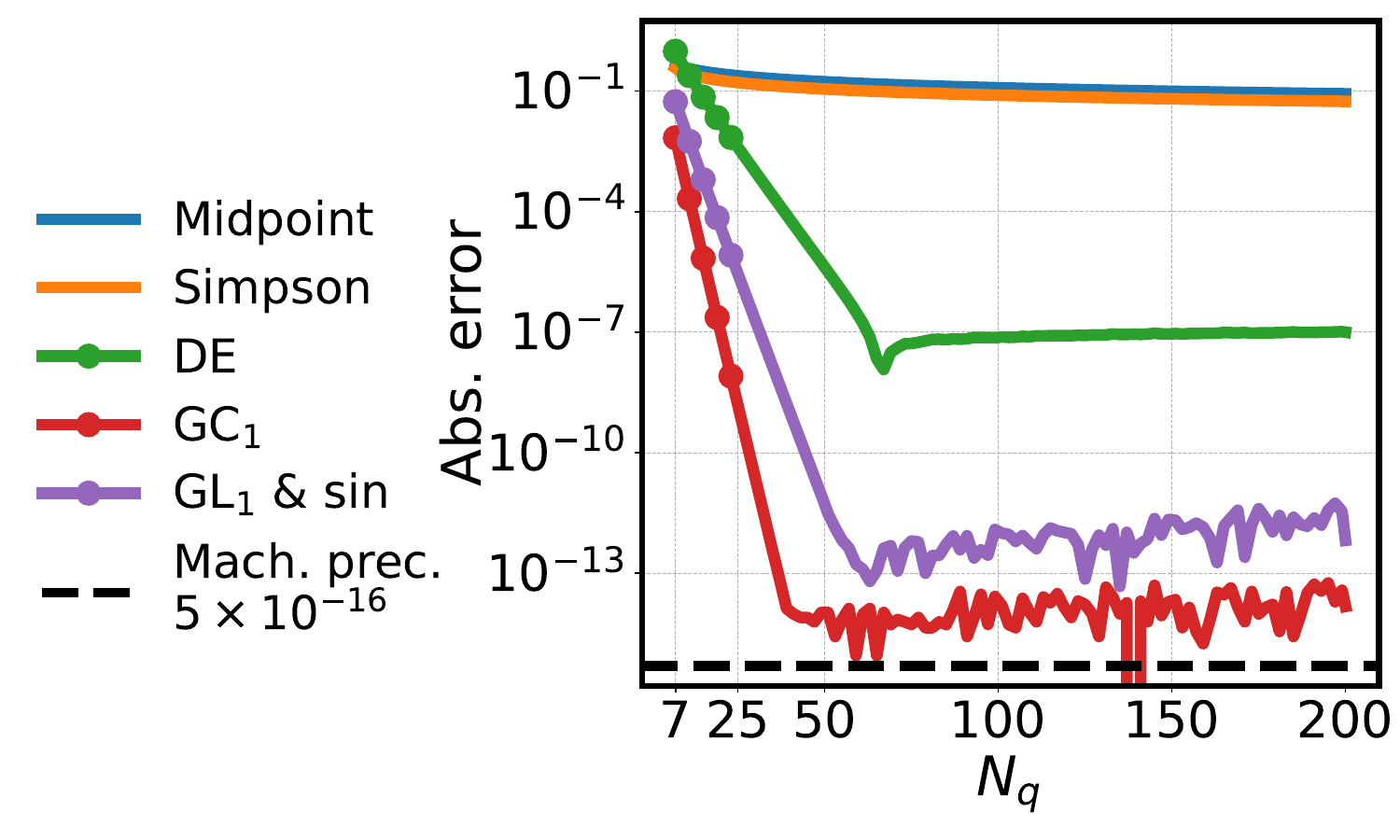}
	\end{subfigure}
	\hfill
	\begin{subfigure}[t]{0.48\textwidth}
		\centering
		\includegraphics[height=3.85cm]{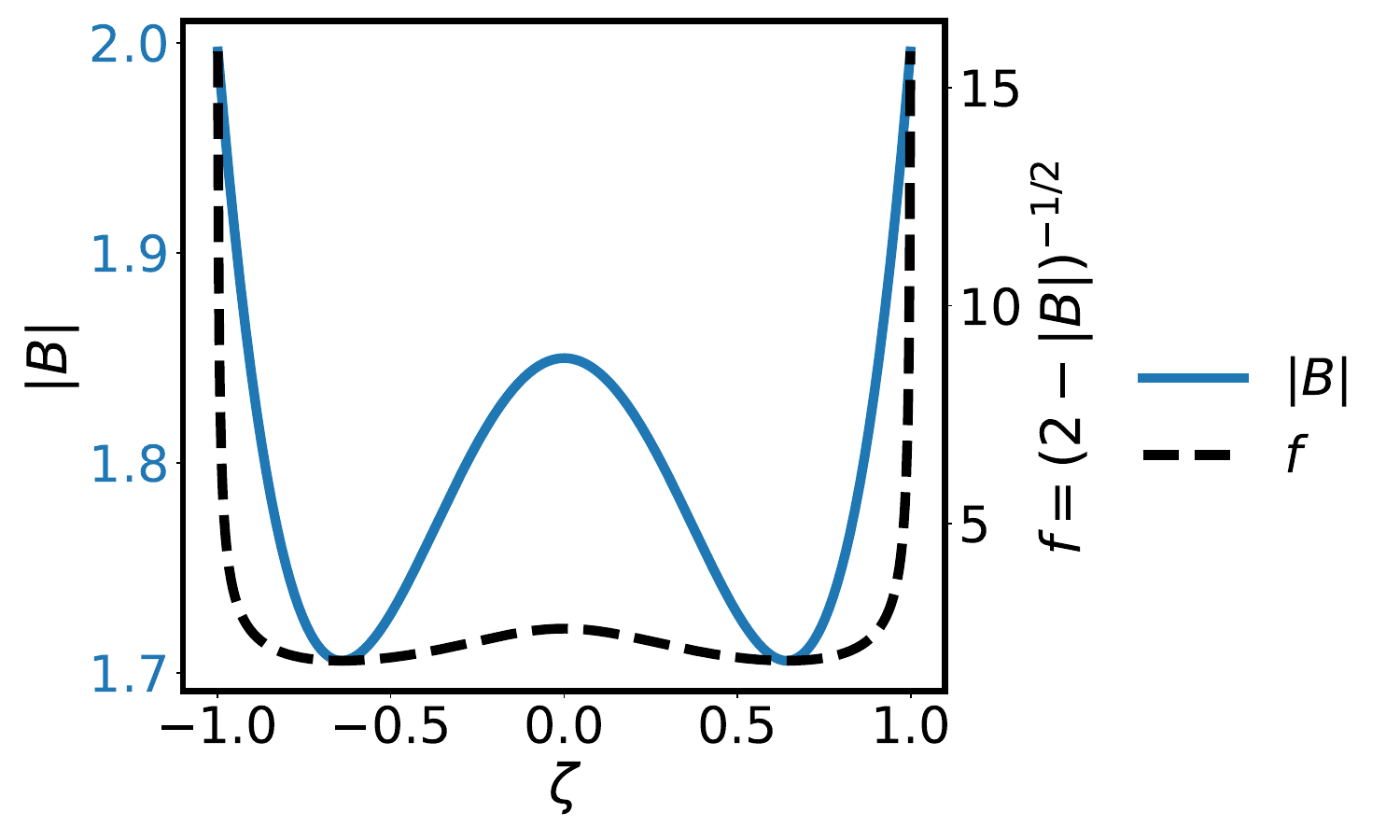}
	\end{subfigure}
	\begin{subfigure}[t]{0.48\textwidth}
		\centering
		\includegraphics[height=3.85cm]{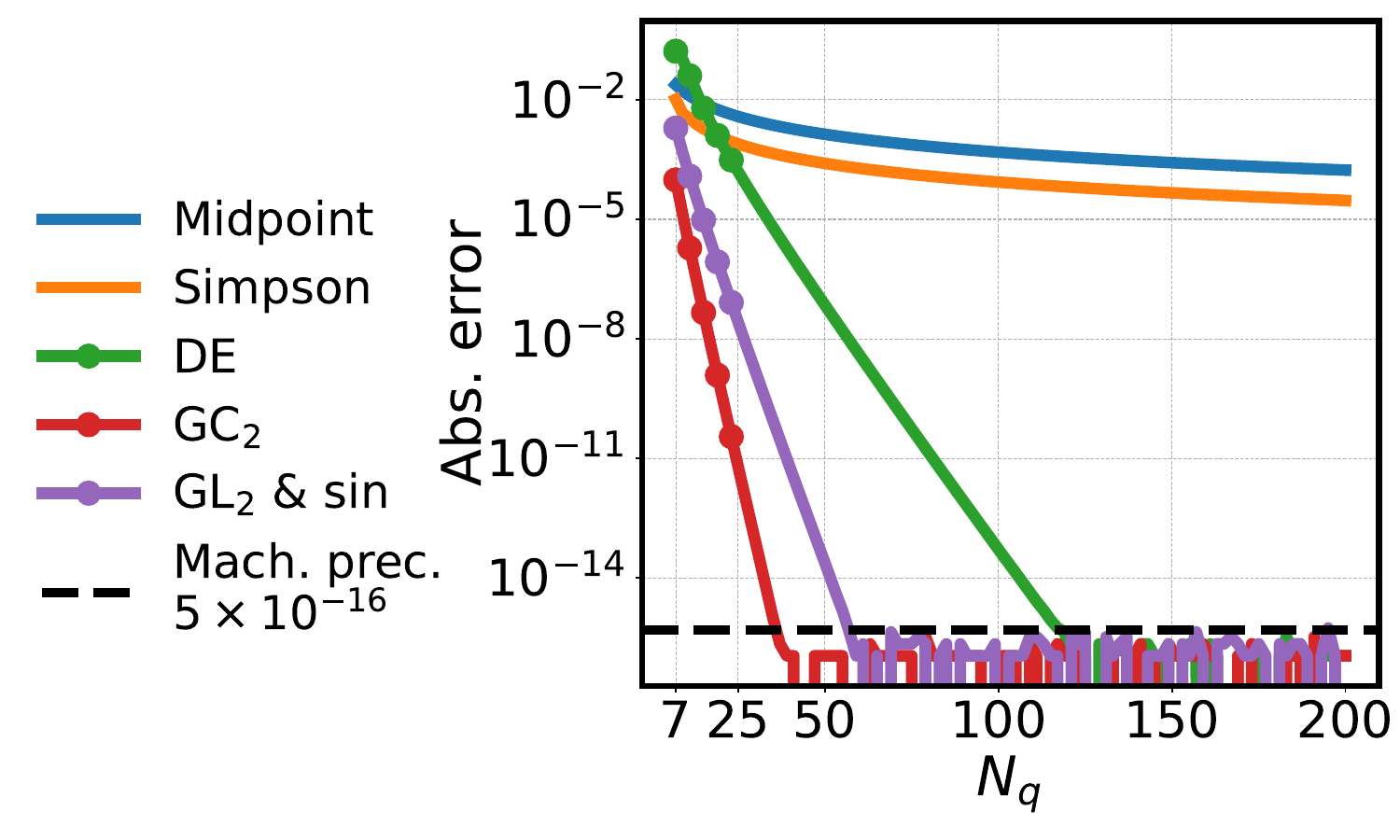}
	\end{subfigure}
	\hfill
	\begin{subfigure}[t]{0.48\textwidth}
		\includegraphics[height=3.85cm]{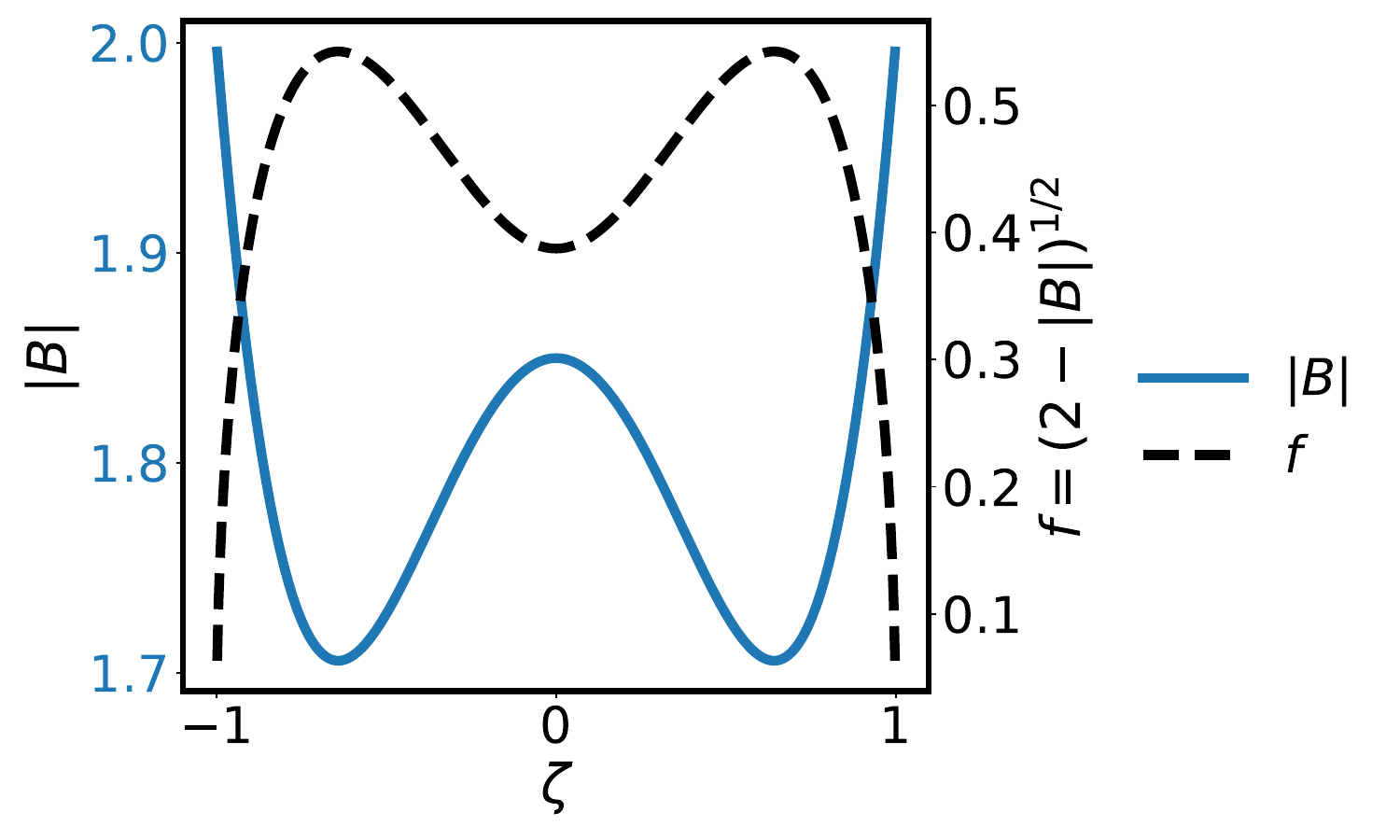}
	\end{subfigure}
	\caption{Convergence of quadratures for the quantity labelled by \(f\).}
	\label{fig:modB-W-shape-shallow}
\end{figure}

\begin{figure}
	\centering
	\begin{subfigure}[t]{0.48\textwidth}
		\centering
		\includegraphics[height=3.85cm]{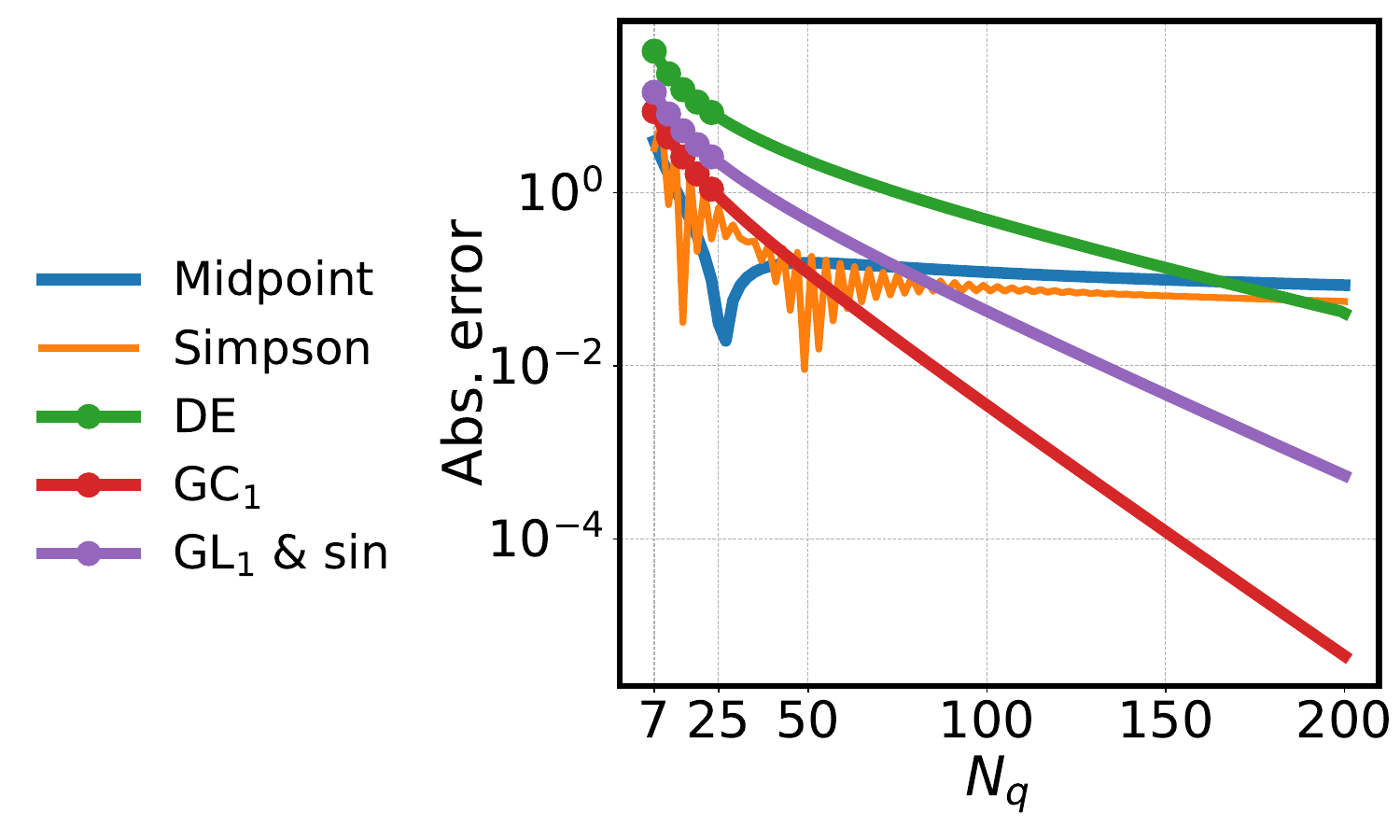}
	\end{subfigure}
	\hfill
	\begin{subfigure}[t]{0.48\textwidth}
		\centering
		\includegraphics[height=3.85cm]{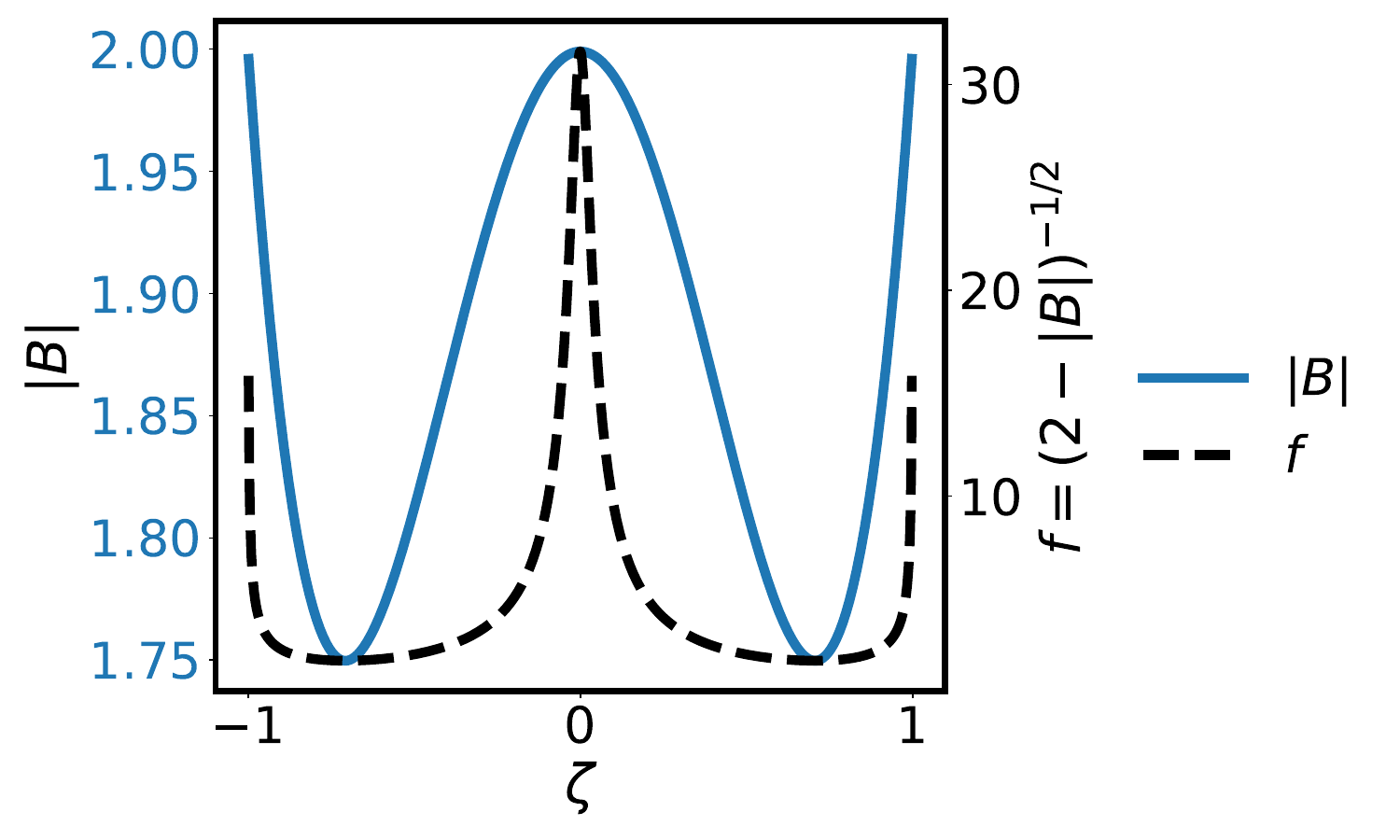}
	\end{subfigure}
	\begin{subfigure}[t]{0.48\textwidth}
		\centering
		\includegraphics[height=3.85cm]{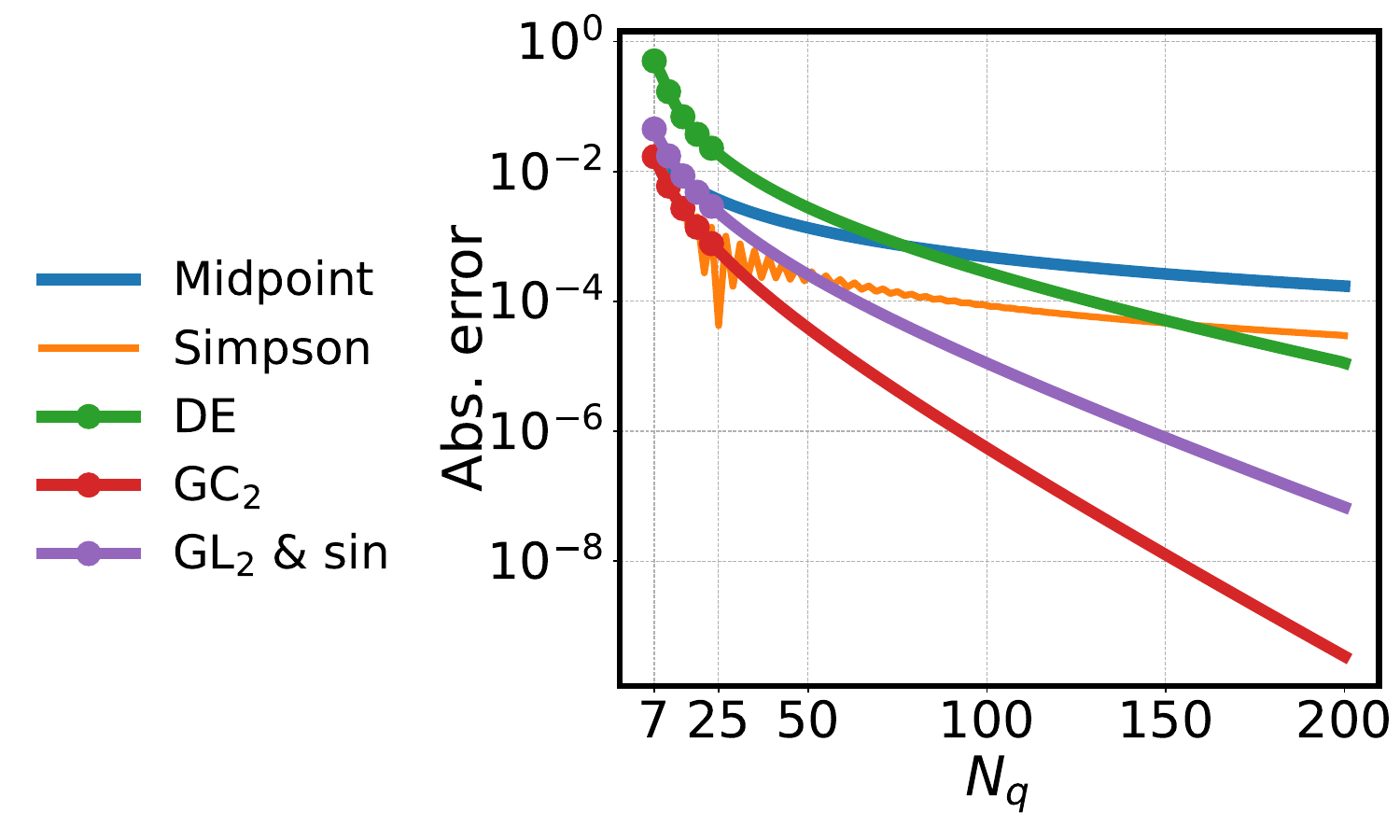}
	\end{subfigure}
	\hfill
	\begin{subfigure}[t]{0.48\textwidth}
		\includegraphics[height=3.85cm]{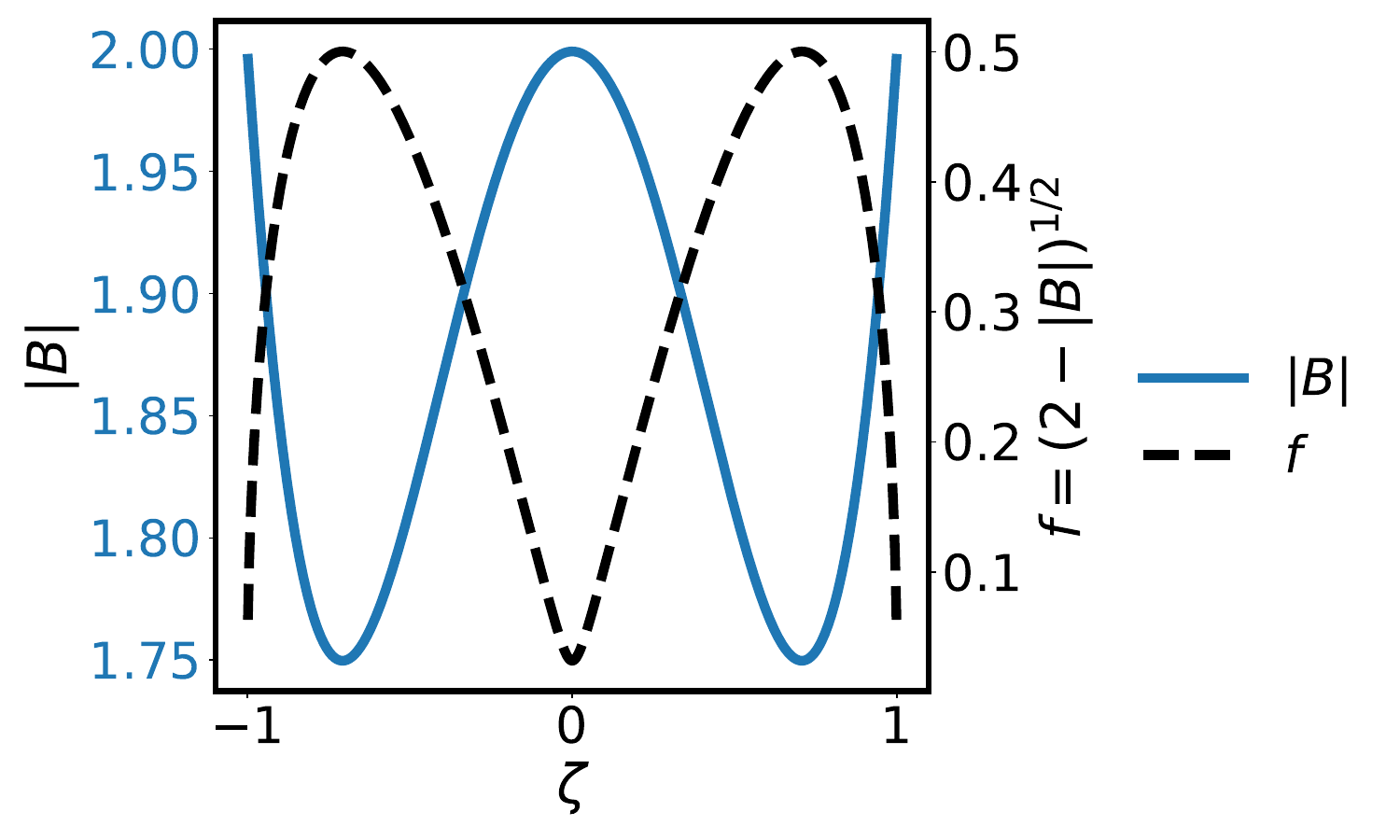}
	\end{subfigure}
	\caption{Convergence of quadratures for the quantity labelled by \(f\). In the top row, the integrand becomes nearly non-integrable as the parallel velocity nearly vanishes at \(\zeta = 0\). In either case, splitting the quadrature there recovers fast convergence.}
	\label{fig:modB-W-shape-deep}
\end{figure}

Our transformation for bounce integrals defines \(z\) such that \(a_1(w, a_2[z]) = \zeta\),
\begin{align}
	a_1 & \colon \begin{cases}
		             \mathbb{N} \times \closeint{-1}{1} \to \mathbb{R}, \\
		             w, z \mapsto (z + 1) [\zeta_{2}(w) - \zeta_{1}(w)]/2 + \zeta_{1}(w),
	             \end{cases}                                          \\
	a_2 & \colon \begin{cases}
		             \closeint{-1}{1} \to \closeint{-1}{1}, \\
		             z \mapsto \sin(\cpi z / 2),
	             \end{cases} \label{eq:gcsin2}                                                                                             \\
	\int_{\zeta_{1}(w)}^{\zeta_{2}(w)} \dl \zeta \, \norm{v_{\parallel}}^{\eta} g(\zeta)
	    & \approx \frac{\zeta_{2}(w) - \zeta_{1}(w)}{2} \sum_{i=1}^{N_q} \sigma_i \norm{v_{\parallel}}^{\eta} g(a_1(w, a_2[z_i])). \label{eq:gcsin}
\end{align}
When neither bounce point is on a local maximum of the potential, the midpoint scheme in \(z\) \eqref{eq:chbgs11} is exponentially accurate,%
\footnote{By the Euler--Maclaurin expansion, the order of convergence is at least the number of continuous derivatives along the field line of \(\norm{B}\) and the surface geometry.
	Their analyticity then implies exponential convergence \citep{10.1137130932132}.
	Although these quantities can be non-analytic \citep{10.10631.872473}, the equilibria produced by \texttt{DESC} yield flux surfaces on which their truncated Fourier series converge fast and these analytic series generate the data.}
\begin{equation}
	\sigma_i = \cpi \sin \groupbrack{\cpi (2 i - 1) / (2 N_q)} / N_q, \qquad a_2[z_i] = \cos \groupbrack{\cpi (2i - 1) /(2 N_q)}.          \label{eq:chbgs11}
\end{equation}
If, in addition, \(\eta = 1\), then \eqref{eq:chbgs22} yields faster convergence,
\begin{equation}
	\sigma_i = \cpi \sin \groupbrack{\cpi i /(N_q + 1)} / (N_q + 1), \qquad a_2[z_i] = \cos \groupbrack{\cpi i /(N_q + 1)}. \label{eq:chbgs22}
\end{equation}
In general, Gauss--Legendre quadrature in \(z\) is exponentially accurate if \eqref{eq:badint} is integrable.
Figures \ref{fig:app-F-quad-compare}, \ref{fig:app-E-quad-compare}, \ref{fig:modB-W-shape-shallow} and \ref{fig:modB-W-shape-deep} illustrate the convergence.

It is often of interest to integrate a nonlinear combination of bounce integrals over \(\varrho\).
Such integrands can be non-smooth in \(\varrho\) due to the logarithmic divergence \citep[\S~4]{Calvo_2017} of \eqref{eq:vdrift} as \(\varrho\) approaches the value of \(\norm{B}\) at any local maxima within or at the bounce points.
The \cite{doi:10.1137/S1064827597325141} quadratures are high-order accurate for these singularities.

We compare the following quadratures in their ability to compute elliptic integrals \eqref{eqn:Elliptic-integral-F}, \eqref{eqn:Elliptic-integral-E}, which are similar to bounce integrals in a simple stellarator geometry.

\begin{enumerate}
	\item Midpoint scheme.
	\item Simpson's 1/3 in the interior completed by a midpoint scheme.
	\item Double exponential (DE) $\tanh-\sinh$.
	\item Implicitly weighted Gauss--Chebyshev of the first ($\mathrm{GC}_1$) \eqref{eq:chbgs11}  or second kind ($\mathrm{GC}_2$) \eqref{eq:chbgs22}.
	\item Gauss--Legendre (\(\mathrm{GL}_1\)) or Gauss--Lobatto--Legendre (\(\mathrm{GL}_2\)) each composed with the $\sin$ transformation in \eqref{eq:gcsin2}.
	      Compared with item (iv), this quadrature offers more resolution near the boundary and less in the interior.
\end{enumerate}
\begin{align}
	\int_{-\arcsin k}^{\arcsin k} \dl \zeta \, (k^2 - \sin^2 \zeta)^{-1/2} & \underset{\protect{\eqref{eqn:I0}}}{=} 2 K(k)
	\label{eqn:Elliptic-integral-F}
	\\
	\int_{-\arcsin k}^{\arcsin k} \dl \zeta \, (k^2 - \sin^2 \zeta)^{1/2}  & \underset{\protect{\eqref{eqn:I1}}}{=} 2 E(k) + 2 (k^2-1) K(k)
	\label{eqn:Elliptic-integral-E}
\end{align}

To further benchmark the quadratures in a magnetic field with ripples, we show two more cases that model particles trapped in the following wells in figures \ref{fig:modB-W-shape-shallow} and \ref{fig:modB-W-shape-deep}.

\subsection{Inverse method} \label{sec:invertalgo}
In this section, we briefly discuss how we find stellarator equilibria.
At static equilibrium, the ideal MHD equations that approximate the behaviour of the plasma reduce to
\begin{align}
	\vec{B} \vecdot \nabla \vec{B} & = \nabla \group{p + \norm{B}^2 / 2},
	\label{eqn:steady-state-ideal-MHD}                                    \\
	\nabla \vecdot \vec{B}         & = 0, \label{eqn:divfree}
\end{align}
which describes a balance between the plasma pressure $p$, magnetic field pressure $\norm{B}^2$ and the effect of field line curvature $\vec{B}\vecdot \nabla \vec{B}$.
We solve the ideal MHD equation using an inverse method.
The computational domain is a solid torus in curvilinear flux coordinates \((\rho, \theta, \zeta)\), where \(\rho = (\psi/\psi_{\text{plasma boundary}})^{1/2}\) and \((\theta, \zeta)\) are angles on a doubly-periodic surface.
Here, \(\Lambda\) and \(\omega\) are to be determined maps that relate the angles \((\theta, \zeta)\) that parametrise a given plasma boundary to the Clebsch angle,
\begin{equation}
	\alpha =  \theta + \Lambda - \iota (\zeta + \omega). \label{eq:alphamap}
\end{equation}
Fourier--Zernike series parametrised in flux coordinates \((\rho, \theta, \zeta)\) are chosen to approximate \(\Lambda\), \(\omega\) and the map to a cylindrical coordinate system \((R, \phi, Z)\) in the lab frame,
\begin{equation}
	R(\rho, \theta, \zeta) \vec{\hat{R}}(\phi) + Z(\rho, \theta, \zeta) \vec{\hat{Z}}. \label{eq:maptorealspace}
\end{equation}
It can be shown from \eqref{eqn:Div-free-B1} that \(\nabla \theta \times \nabla \zeta \neq \vec{0}\) implies
\begin{align}
	\vec{B} \vecdot \nabla \theta & = -[\nabla \psi \vecdot (\nabla \theta \times \nabla \zeta)] \group*{\diffp{\alpha}{\zeta}}_{\psi, \theta}, \label{eq:cond0}  \\
	\vec{B} \vecdot \nabla \zeta  & = + [\nabla \psi \vecdot (\nabla \theta \times \nabla \zeta)] \group*{\diffp{\alpha}{\theta}}_{\psi, \zeta}. \label{eq:cond1}
\end{align}
Thus, we find equilibria by searching for a combination \((R, Z, \Lambda, \omega)\) to reduce the force balance error \eqref{eqn:steady-state-ideal-MHD} at a set of collocation points using pseudo-spectral methods, subject to constraints on the pressure profile and the rotational transform or toroidal current profile.
This boundary value problem is then solved as a minimisation problem using a trust-region method.
In an optimisation constrained by force balance, varying \((R, Z, \Lambda, \omega)\) changes the magnetic field and \eqref{eq:maptorealspace} such that \eqref{eq:alphamap} remains valid.

Two advantages of this inverse approach for optimisation with bounce-averaged objectives are stated as follows.
\begin{enumerate}
	\item The variables \((\theta, \zeta)\) on the boundary surface may be constructed so that maps parametrised in these coordinates are spectrally condensed  \citep{speccondense, Hindenlang_2025}. Consequently, maps parametrised in \((\rho, \theta, \zeta)\) in the plasma volume tend to have spectral expansions that converge more rapidly.
	\item Force balance and other geometric objectives are best computed on a particular grid in \((\rho, \theta, \zeta)\), which is fixed throughout optimisation.
	      This ensures the spectral basis for \((R, Z, \Lambda, \omega)\) can be precomputed, avoiding \enquote{off-grid} interpolation of a three-dimensional basis that bottlenecks pseudo-spectral codes \citep[\S~10.7]{boyd2013chebyshev}.
	      Furthermore, if the coordinate system varies throughout the optimisation, then so does the optimal grid for interpolation and quadrature.
	      To preserve spectral accuracy, a pseudo-spectral code must first find this optimal grid and compute the basis there.
	      This \enquote{moving-grid} interpolation is doubly expensive in optimisation because the mentioned tasks must also be differentiated, which consumes significant memory.
\end{enumerate}

These qualities enable faster generation of magnetic field data, which we discuss in the following section.

\subsection{Fast interpolation} \label{sec:mitoff}
In this section, we outline our method for fast interpolation.

The Zernike basis concentrates the frequency transform of smooth maps on discs at lower frequencies than geometry-agnostic tensor-product bases.
Boyd \& Yu (\citeyear{BOYD20111408}) show the required number of spectral coefficients is typically half that of Fourier--Chebyshev.
This ensures an optimisation that varies a finite number of coefficients in the Fourier--Zernike series for \((R, Z, \Lambda, \omega)\) at a time has more freedom compared with expansions in other bases.
However, the Zernike basis is expensive to evaluate.

Our algorithm computes the Fourier--Zernike basis once prior to optimisation on a uniform \(K_{\theta} \times K_{\zeta}\) grid in \((\theta, N_{\text{FP}} \zeta) \in \clopenint{0}{2 \cpi}^2\) on each surface.
Any smooth, periodic map \(g\) required by the objective is computed from \((R, Z, \Lambda, \omega)\) on this grid and interpolated with a fast Fourier transform (FFT).
The resulting Fourier series are evaluated using type-2 non-uniform FFTs with computational cost that is linearithmic in \(K_{\theta} K_{\zeta}\) plus linear in the number of points to evaluate \citep{doi:10.1137/18M120885X, BARNETT20211, shih2021cufinufftloadbalancedgpulibrary, unalmis2025jaxfinufft},
\begin{align}
	g_{k_{\theta} k_{\zeta}} & = \frac{c_{k_{\theta}}}{4 \cpi^2} \iint_{\clopenint{0}{2\cpi}^2} \dl \theta \, \dl (N_{\text{FP}} \zeta) \; g(\theta, N_{\text{FP}} \zeta) \e^{-\im k_{\theta} \theta} \e^{-\im k_{\zeta} N_{\text{FP}} \zeta}, \label{eq:sp}                               \\
	g(\alpha, \zeta)         & = \sum_{k_{\theta}=0}^{\floor{K_{\theta}/2}} \sum_{k_{\zeta}=-\floor{K_{\zeta} /2}}^{\ceil{K_{\zeta}/2} - 1} \text{Real} \group*{g_{k_{\theta} k_{\zeta}} \e^{\im k_{\theta} \theta(\alpha, \zeta)} \e^{\im k_{\zeta} N_{\text{FP}} \zeta}} \label{eq:sp2}, \\
	c_{k_{\theta}}           & = 1 \text{ if } k_{\theta} \in \set{0, K_{\theta}/2} \text{ else } 2. \nonumber
\end{align}

\subsection{Map to the mesh of field lines} \label{sec:movingmap}
\begin{figure}
	\centering
	\begin{subfigure}[t]{0.48\textwidth}
		\centering
		\includegraphics[height=4.2cm]
		{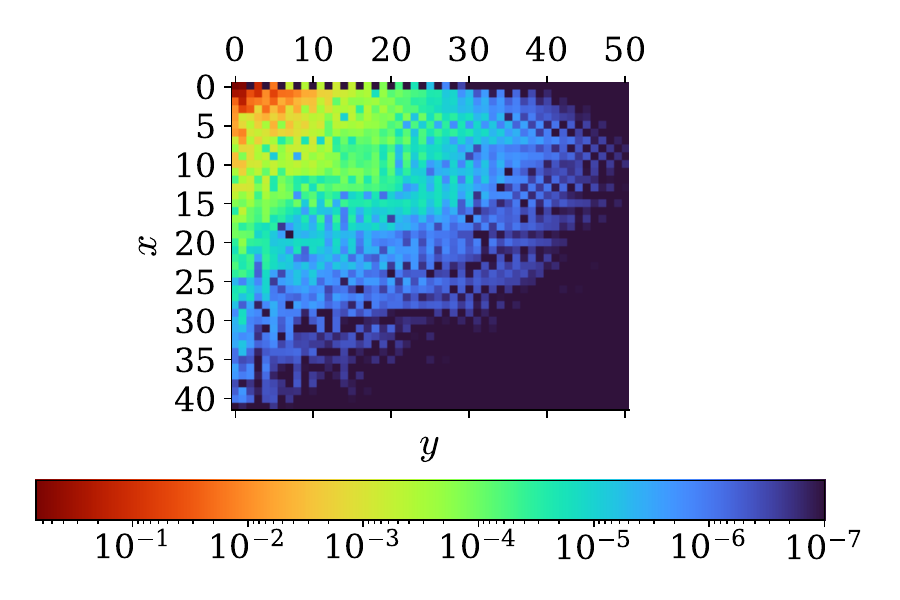}
		\caption{\(\norm{a_{xy}}\) on the plasma boundary of an NCSX stellarator with \(N_\text{FP} = 3\).}
		\label{fig:subfiga}
	\end{subfigure}
	\hfill
	\begin{subfigure}[t]{0.48\textwidth}
		\centering
		\includegraphics[height=4.2cm]{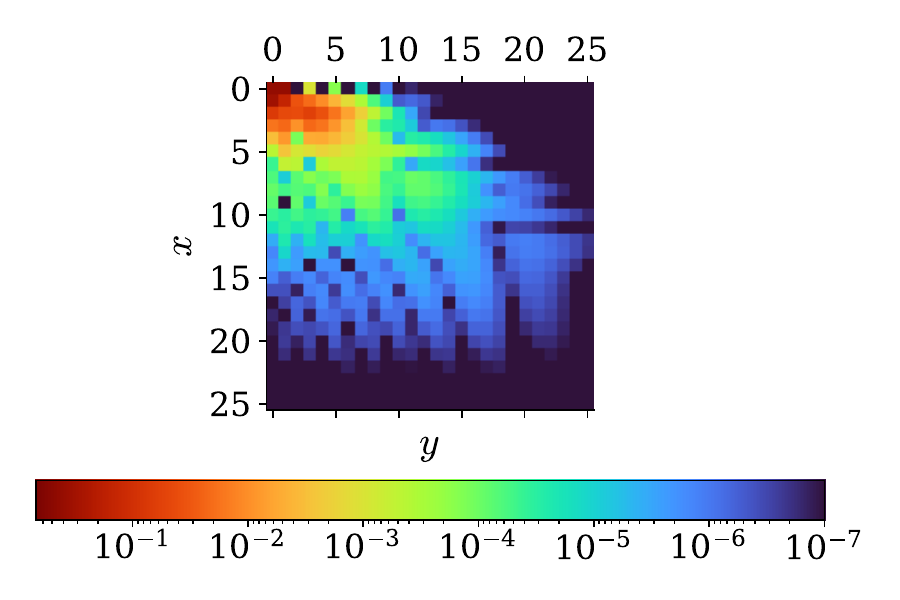}
		\caption{\(\norm{a_{xy}}\) on the plasma boundary of a Heliotron stellarator with \(N_\text{FP} = 19\).}
		\label{fig:subfigb}
	\end{subfigure}
	\caption{Convergence of the spectral projection of \(\alpha, \zeta \mapsto \theta - \alpha\) onto the Fourier--Chebyshev basis \eqref{eq:thetaseries}.
		Equation \protect{\eqref{eq:alphamap}} was solved to error \(\leq 10^{-7}\).
		Note that if \(\omega \to \Lambda / \iota\), then \(\theta - \alpha \to \iota \zeta\), so the spectral width reduces to one parameter. Thus, if the optimiser is motivated to match higher frequency spectral coefficients of \(\omega\) with \(\Lambda / \iota\), then field lines can be tracked at lower resolution.}
	\label{fig:streammapconvergence}
\end{figure}

Evaluating maps along field lines requires finding the position of the field lines on some grid.
To identify the coordinate \(\theta\) at a given point \((\alpha, \zeta)\) one may solve \eqref{eq:alphamap}.
To avoid repeating that inversion everywhere our objective demands, we compute the spectral projection \(\set{a_{xy}}\) of the map \(\alpha, \zeta \mapsto \theta - \alpha\) onto a tensor-product basis \(\set{b_{x y}}\) that is orthogonal with respect to some weight \(\varsigma\),
\begin{equation}
	a_{x y} \sim \iint \dl \alpha \, \dl \zeta \; (\theta - \alpha) \, \varsigma b^{*}_{x y}(\alpha, \zeta). \label{eq:coefspec}
\end{equation}
The Fourier--Chebyshev basis defined on the field period \((\alpha, N_\text{FP} \zeta) \in \clopenint{0}{2 \cpi}^2\) is chosen for reasons discussed by \citet[\S\S~5.5, 5.6, 6.3.4]{Mason2002Chebyshev} and \citet[\S~4.5]{boyd2013chebyshev},
\begin{equation}
	b_{xy}(\alpha, \zeta) = \e^{\im x \alpha} \cos (y \arccos [N_\text{FP} \zeta / \cpi - 1]).
\end{equation}
On each flux surface, \eqref{eq:alphamap} is solved on a tensor-product grid of size \(X \times Y\) on the Fourier nodes across field lines and the Chebyshev nodes along field lines using Newton iteration with a backtracking line search.
The series \eqref{eq:coefspec} is computed by interpolating \(\theta - \alpha\) on that grid with a discrete cosine transform along field lines, followed by an FFT across field lines.
The convergence of the series is illustrated in figure \ref{fig:streammapconvergence}.

\begin{figure}
	\centering
	\begin{subfigure}{\textwidth}
		\centering
		\hspace{-0.5cm}
		\includegraphics[width=0.85\linewidth]{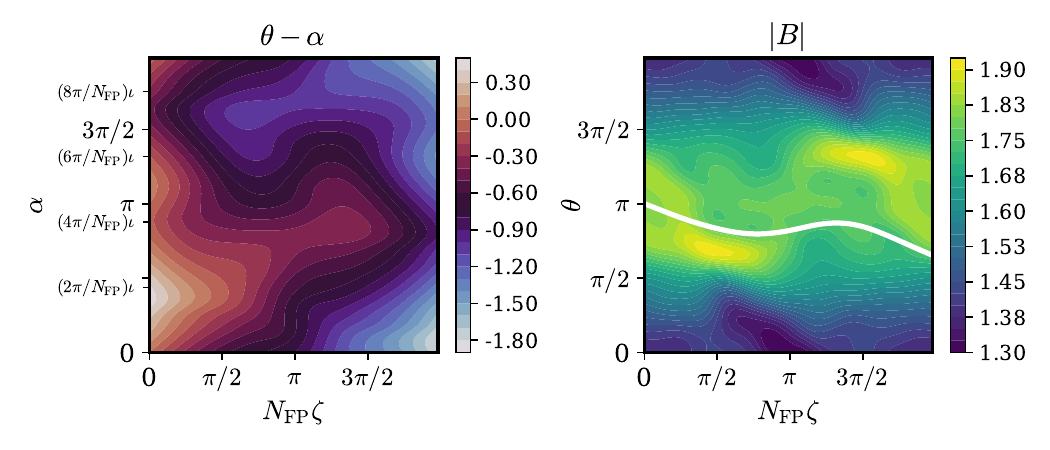}
		\vspace{-0.2cm}
		\caption*{\(\alpha, \zeta \mapsto \theta - \alpha\) is shown on the left. On the right, \(\alpha = \cpi\) is shown in white.}
	\end{subfigure}
	\begin{subfigure}{0.55\textwidth}
		\centering
		\hspace{-1cm}
		\includegraphics[width=\linewidth]{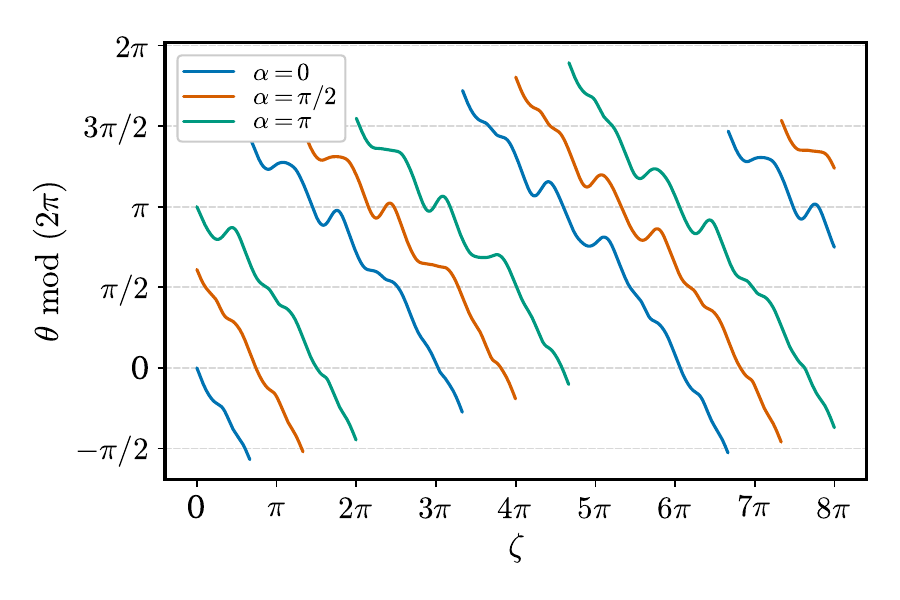}
		\vspace{-0.3cm}
	\end{subfigure}
	\caption{\(\theta\) on the plasma boundary of an NCSX stellarator with \(N_{\text{FP}} = 3\).}
	\label{fig:llll}
\end{figure}

To extend the map beyond a single field period, we use
\begin{align}
	\theta                & \equiv \alpha_{\text{mod}} + \sum_{x=0}^{\floor{X/2}} \sum_{y=0}^{Y - 1} \text{Real} \group*{a_{xy} b_{xy}(\alpha_{\text{mod}}, \zeta_{\text{mod}})} \pmod{2\cpi}, \label{eq:thetaseries} \\
	\alpha_{\text{mod}}   & = \alpha_{\text{shift}} \bmod (2 \cpi), \nonumber                                                                                                                                         \\
	\alpha_{\text{shift}} & = \alpha + \iota \floor{N_{\text{FP}} \zeta / (2 \cpi)} 2 \cpi /N_{\text{FP}}, \nonumber                                                                                                  \\
	\zeta_{\text{mod}}    & = \zeta \bmod (2 \cpi / N_{\text{FP}}). \nonumber
\end{align}
The equivalence \eqref{eq:thetaseries} is due to \(\alpha + \iota \zeta = \alpha_{\text{shift}} + \iota \zeta_{\text{mod}}\) and uniqueness of solutions to \eqref{eq:alphamap}.
Figure \ref{fig:llll} shows an illustration.
The construction and evaluation of this series is accelerated with partial summation \citep[\S~10]{boyd2013chebyshev}.

This approach avoids issues that result from changing the basis for \((R, Z, \Lambda, \omega)\) at each optimisation step (Appendix \ref{specbasistopest}).

\vspace{-0.5cm}

\subsection{Jacobian of the map to the mesh of field lines} \label{sec:jacmap}
In this section, we explain how we accelerate the iterative solve discussed in the previous section throughout optimisation.
To bypass differentiating the iterative solve, we write the tangent and adjoint methods directly \citep[\S\S~3.3.3, 3.9.2]{sapienza2025differentiableprogrammingdifferentialequations}.
For this task, we leverage the implicit function theorem to differentiate solutions \(\theta^\star\) to \eqref{eq:alphamap} with respect to the optimisable parameters, denoted here with \(\vec{x}_{\text{opt}}\).
Define
\begin{equation}
	f \colon \vec{x}_{\text{opt}}, \theta \mapsto \theta + \Lambda - \iota (\zeta + \omega) - \alpha.
\end{equation}
Let \((\vec{x}_{\text{opt}}^\star, \theta^\star)\) satisfy \(f(\vec{x}_{\text{opt}}^\star, \theta^\star) = 0\),
\begin{equation}
	\diffp{f}{\theta}(\vec{x}_{\text{opt}}^\star, \theta^\star) =  1 + \diffp{(\Lambda - \iota \omega)}{\theta}(\vec{x}_{\text{opt}}^\star, \theta^\star) \underset{\protect{\eqref{eq:alphamap}}}{=} \group*{\diffp{\alpha}{\theta}}_{\psi, \zeta}(\vec{x}_{\text{opt}}^\star, \theta^\star). \label{eq:lastalmost}
\end{equation}
In the \((\psi, \alpha, \zeta)\) covariant basis, the only non-zero component of the non-vanishing magnetic field is \eqref{eq:cond1}, so the derivative \eqref{eq:lastalmost} is invertible.
By the implicit function theorem, \(\theta^\star\) is a continuously differentiable map of \(\vec{x}_{\text{opt}}\) and \(f(\vec{x}_{\text{opt}}, \theta^\star(\vec{x}_{\text{opt}})) = 0\) near \(\vec{x}_{\text{opt}}^\star\).
Moreover,
\begin{equation}
	\diffp{\theta^\star}{\vec{x}_{\text{opt}}}(\vec{x}_{\text{opt}}) = -\groupbrack*{\diffp{f}{\theta}(\vec{x}_{\text{opt}}, \theta^\star(\vec{x}_{\text{opt}}))}^{-1} \diffp{f}{\vec{x}_{\text{opt}}}(\vec{x}_{\text{opt}}, \theta^\star(\vec{x}_{\text{opt}})). \label{eq:finallylast}
\end{equation}
Thus, we differentiate directly through the solution \(\theta^\star\).
Likewise, after updating \(\vec{x}_{\text{opt}}\), we use \eqref{eq:finallylast} to warm start the next Newton iteration at an initial value that is correct to first order.

%% file: 4_optimization.tex
\section{Optimisation for reduced neoclassical transport}
\label{sec:optimization}
\begin{figure}
	\centering
	\begin{subfigure}[t]{0.28\textwidth}
		\raggedleft
		\includegraphics[height=2.5cm, trim={0, 1, 0, 1cm}, clip]{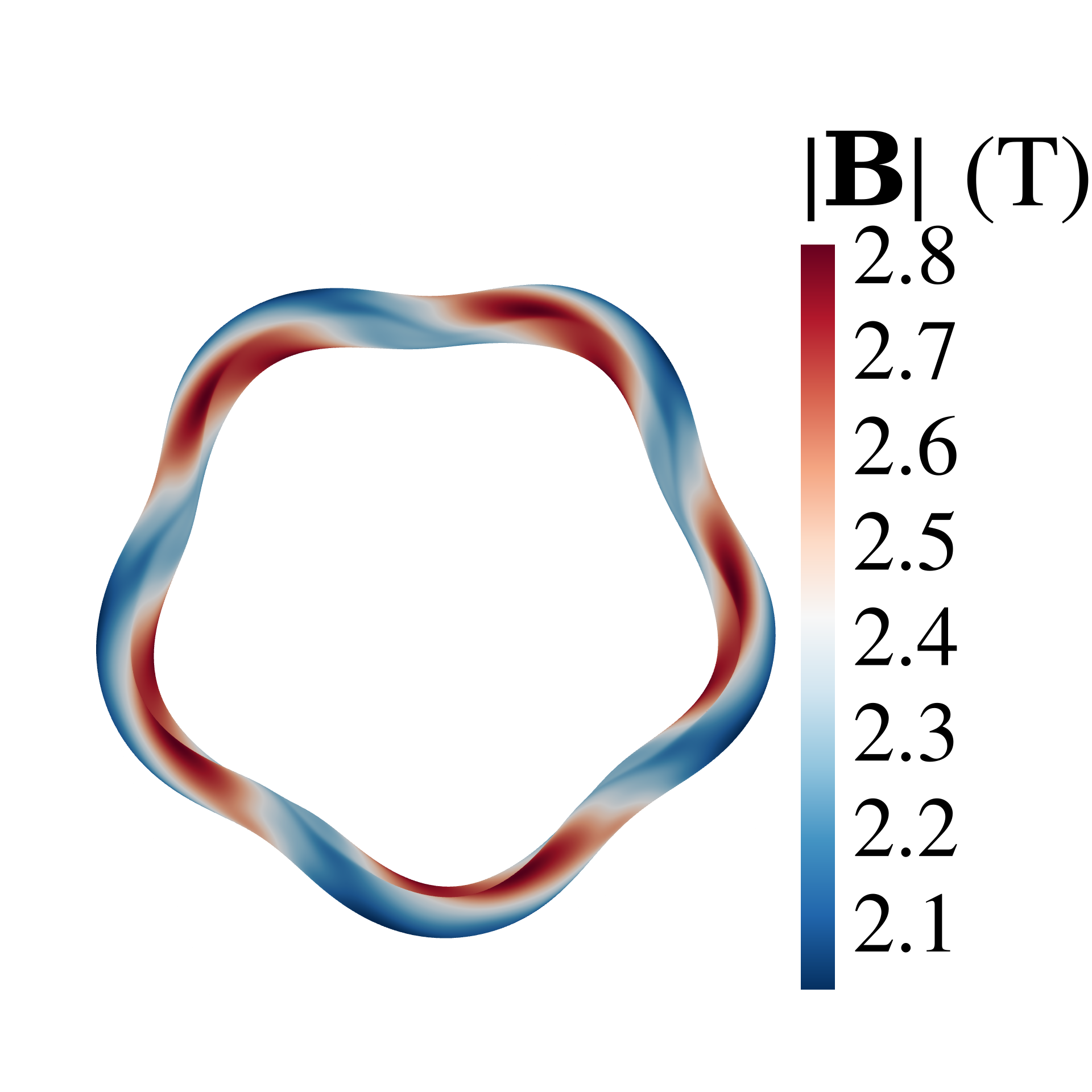}
	\end{subfigure}
	\begin{subfigure}[t]{0.4\textwidth}
		\centering
		\includegraphics[height=3.75cm]{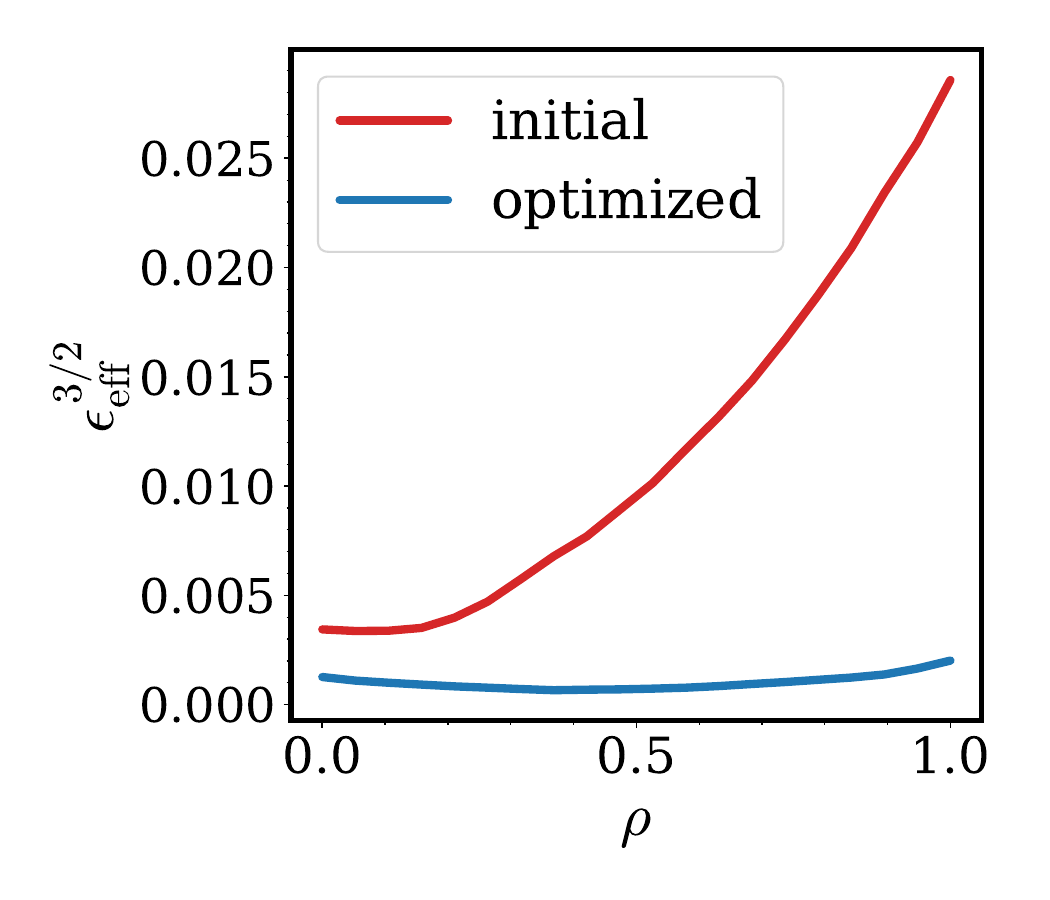}
		\caption{Neoclassical transport optimisation.}
	\end{subfigure}
	\begin{subfigure}[t]{0.28\textwidth}
		\raggedright
		\includegraphics[height=2.5cm, trim={0, 1cm, 0, 1cm}, clip]{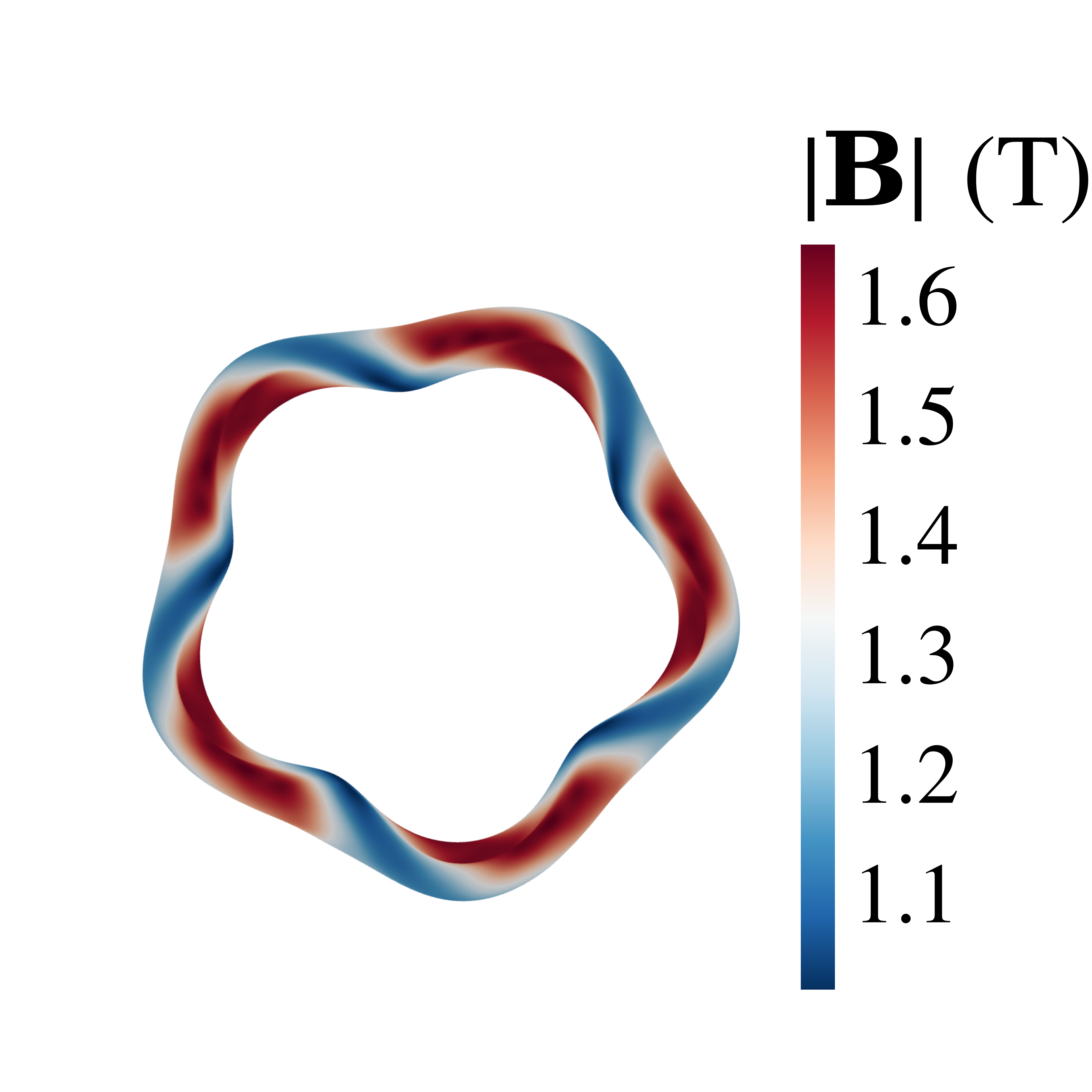}
	\end{subfigure}

	\begin{subfigure}[b]{0.48\textwidth}
		\centering
		\includegraphics[width=0.8\linewidth]{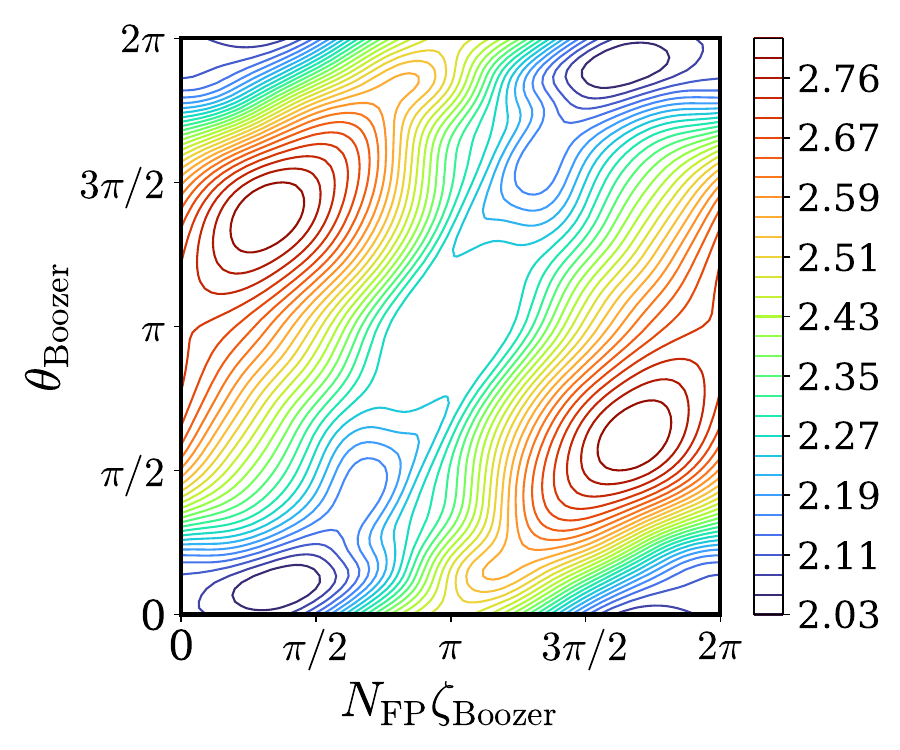}
		\caption{Initial equilibrium \(\norm{B}\).}
		\label{figinitbo}
	\end{subfigure}
	\begin{subfigure}[b]{0.48\textwidth}
		\centering
		\includegraphics[width=0.8\linewidth]{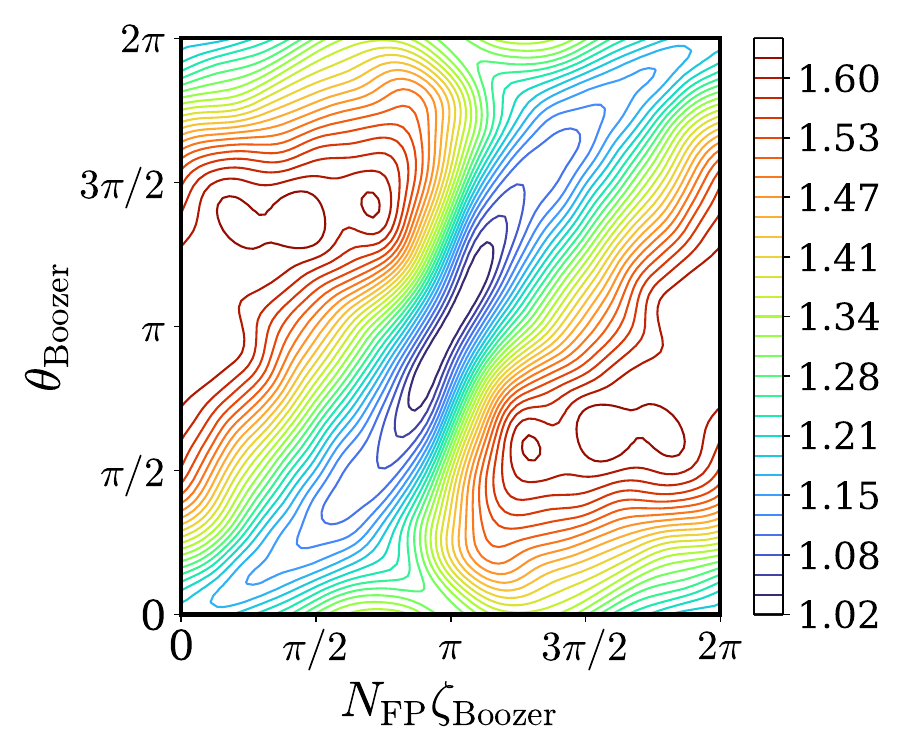}
		\caption{Optimised equilibrium \(\norm{B}\).}
		\label{figoptbo}
	\end{subfigure}

	\begin{subfigure}[b]{0.48\textwidth}
		\centering
		\includegraphics[width=0.8\linewidth]{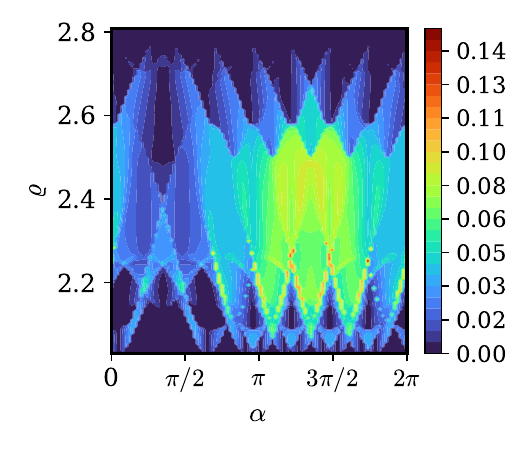}
		\caption{Initial bounce-averaged radial drifts.}
		\label{figgam1}
	\end{subfigure}
	\begin{subfigure}[b]{0.48\textwidth}
		\centering
		\includegraphics[width=0.8\linewidth]{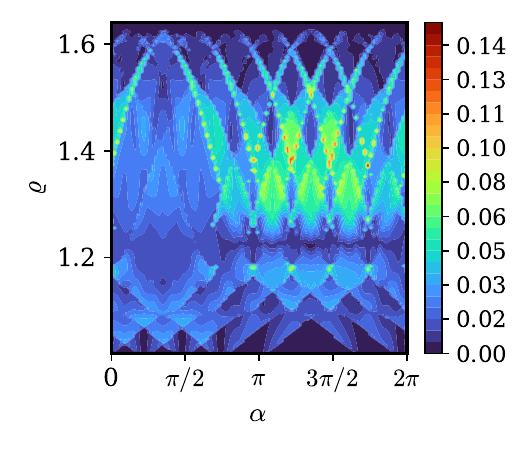}
		\caption{Optimised bounce-averaged radial drifts.}
		\label{figgam2}
	\end{subfigure}

	\begin{subfigure}[t]{0.9\textwidth}
		\centering
		\includegraphics[width=0.82\linewidth]{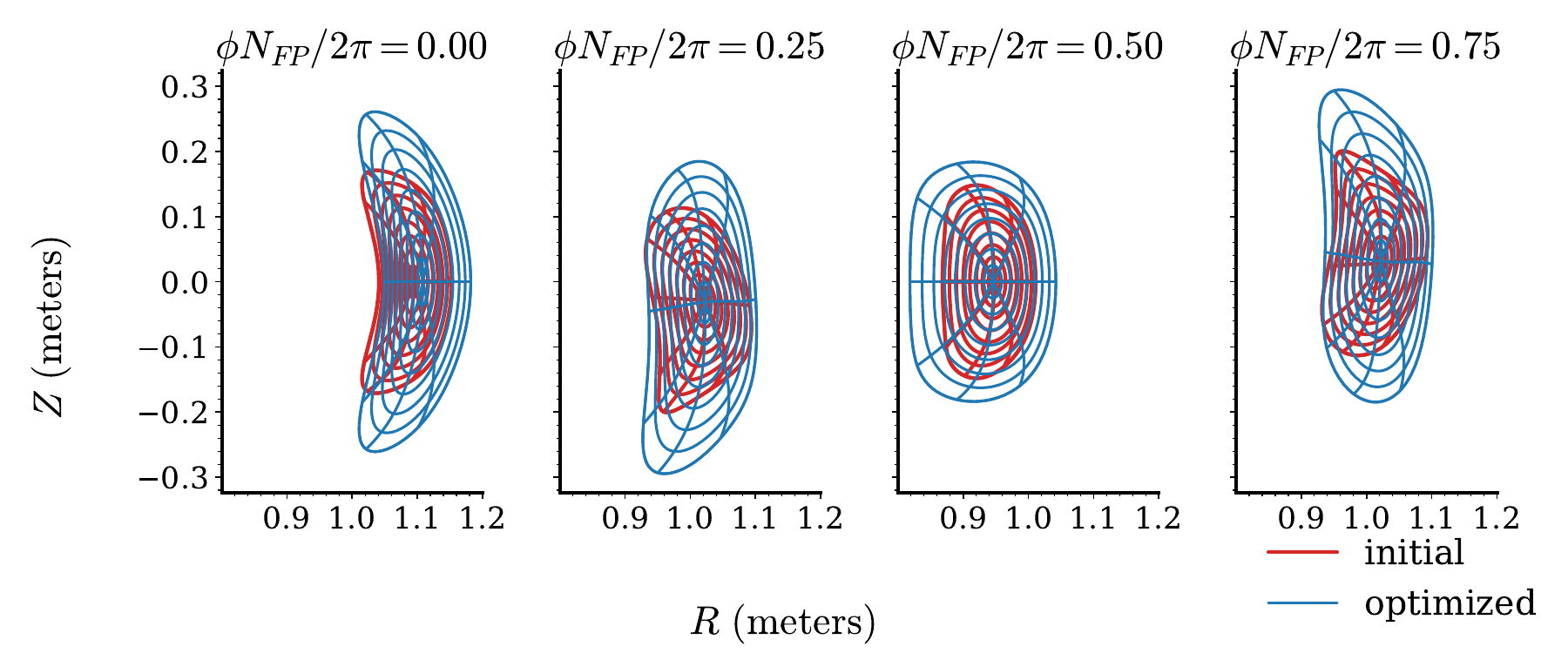}
		\caption{Cross sections are compared at constant toroidal angles in the lab frame.}
	\end{subfigure}

	\caption{An OH transport optimisation. Panels (\protect{\subref{figinitbo}}) and (\protect{\subref{figoptbo}}) are shown in Boozer coordinates \protect{\citep{d2012flux}}. Panels (\protect{\subref{figgam1}}) and (\protect{\subref{figgam2}}) show bounce-averaged radial drifts, summed in magnitude over \(\zeta \in \openint{0}{2\cpi}\). The size of the region with large drifts appears reduced.}
	\label{fig:OH-outputs}
\end{figure}

We present an optimisation starting from a finite-beta helically omnigenous (OH) equilibrium.
Finite-beta refers to the non-zero ratio of plasma pressure to magnetic pressure.
We target flux surfaces near the boundary to reduce the effective ripple while maintaining reasonable elongation and curvature.
With weights, $w_{\mathrm{A}}, w_{\mathrm{C}}, w_{\mathrm{E}}, w_{\mathrm{O}}, w_{\mathrm{R}}$, the
objective \eqref{eq:obj} is minimised while ensuring ideal MHD force balance \eqref{eqn:steady-state-ideal-MHD} is maintained,
\begin{equation}
	w_{\mathrm{A}} f_{\mathrm{aspect}}^2 + w_{\mathrm{C}} f_{\mathrm{curv}}^2 + w_{\mathrm{E}} f_{\mathrm{elongation}}^2 + w_{\mathrm{O}} f_{\mathrm{omni}}^2 + w_{\mathrm{R}} f_{\mathrm{ripple}}^2. \label{eq:obj}
\end{equation}
The initial equilibrium along with the definitions of the curvature and elongation objectives are provided by~\cite{Gaur_2025} and~\cite{gaur_2024_13887566}.
The results are presented in figure \ref{fig:OH-outputs}.
The optimisation took less than two hours with a GPU~\citep{nvidia_a100_gpu}.

The omnigeneity objective is based on the work of~\cite{dudt2024magnetic}, where it was shown that optimising for omnigeneity can in turn reduce the effective ripple.
Directly optimising to reduce the effective ripple instead has the advantage that the optimiser is not biased towards a user-specified omnigenous field.
For example, in \cite{Gaur_Panici_Elder_Landreman_Unalmis_Elmacioglu_Dudt_Conlin_Kolemen_2025}, we used this property to optimise for an umbilic boundary, while maintaining a low effective ripple without biasing the optimiser towards an omnigenous field with a specific helicity.

It should be noted that the assumptions used to derive the effective ripple increase in validity as the magnetic field becomes more omnigenous.
For example, bounce-averaging assumes the radial orbit width is small compared with the magnetic field gradient scale length \(\Delta r \ll L_{B}\). When finite orbit width effects become dominant \citep{Herbemont_Parra_Calvo_Velasco_2022}, particles may traverse \enquote{potato} orbits requiring a more global treatment \citep{10.1063/1.1499952}.
Hence, there is utility in optimisation that uses both objectives, either simultaneously or in succession.

%% file: 5_summary.tex
\section{Conclusions}
\label{sec:conclusion}
In this work, we optimised a finite-beta configuration to directly reduce neoclassical transport using reverse-mode differentiation.
More generally, we upgraded the \texttt{DESC} stellarator optimisation suite for fast, accurate, automatically differentiable bounce-averaging.
We discussed how we perform moving-grid interpolation without sacrificing spectral accuracy.
This accuracy ensures that changes in the objective due to small changes in controllable parameters reflect genuine improvement or degradation rather than noise due to error.
Therefore, optimisation is more likely to be successful.

Our algorithm enables optimisation for many objectives to improve stellarator performance.
These include maximisation of the second adiabatic invariant (\citealp[\S~3.7]{helander2014theory}; \citealp{Rodríguez_Helander_Goodman_2024}), energetic particle confinement \citep{10.1063/1.2912456, velasco2021model} and proxies for gyrokinetic turbulence such as the available energy~\citep{mackenbach2022available, Mackenbach_Proll_Wakelkamp_Helander_2023}.
We have added all these objectives to \texttt{DESC}.
Some of these objectives have previously had limited use in optimisation due to expensive computational requirements or difficulty finding desirable configurations.
Further demonstration of optimisation with them remains as future work.

\bookmarksection{Acknowledgements}
\textit{Editor Per Helander thanks the referees for their advice in evaluating this article.}

\vspace{-0.2cm}
\bookmarksection{Funding}
This work is funded through the SciDAC program by the US Department of Energy, Office of Fusion Energy Science, and Office of Advanced Scientific Computing Research under contract numbers DE-AC02-09CH11466, DE-SC0022005, and Field Work Proposal No. 1019.
This work was also funded by the Peter B. Lewis Fund for Student Innovation in Energy and the Environment.
This research used the resources of the Della computing cluster at Princeton University.

\vspace{-0.2cm}
\bookmarksection{Declaration of interests}
The authors declare no competing interests.

\vspace{-0.2cm}
\bookmarksection{Author Contributions}
\orcidlink{0000-0002-1755-6764} Kaya Unalmis: Conceptualization (lead); Methodology (lead); Formal analysis (lead); Investigation (lead); Software (lead); Validation (lead); Visualization (lead); Data curation (lead); Writing -- original draft (lead); Writing -- review \& editing (lead).

\orcidlink{0000-0003-4367-9052} Rahul Gaur: Software; Writing -- review \& editing.

\orcidlink{0000-0001-8366-2111} Rory Conlin: Methodology; Software.

\orcidlink{0000-0003-0736-4360} Dario Panici: Writing -- review \& editing.

\orcidlink{0000-0003-4212-3247} Egemen Kolemen: Supervision; Writing -- review \& editing; Funding acquisition; Project administration.

%% file: A_open.tex
\section{Open source code} \label{app:a}
The implementation and supplementary information describing how correctness of automatic differentiation is enforced may be found in the \texttt{DESC} repository \citep{desc_git}.
The implementation uses accelerated linear algebra \texttt{XLA} and Google's \texttt{JAX} library \citep{jax2018github}.
\texttt{JIT} compilation in \texttt{JAX} compiles the code at the start of an optimisation.
Optimisation may be accelerated on CPUs, GPUs and TPUs.

%% file: B_ripple.tex
\section{Effective ripple} \label{sec:neoderivation}
We explain a short derivation of \(\epsilon_{\text{eff}}\), similar to the one used by~\cite{nemov1999evaluation}.

To obtain an explicit expression for \(\Gamma\), we will bounce-integrate the drift-kinetic equation \eqref{eqn:Neoclassical-equation}.
Applying this operator to the drift-kinetic equation discretises the spatial coordinate \(\zeta\) to a set of integral equations labelled by the magnetic well index \(w\).

The collision operator in \eqref{eqn:Neoclassical-equation} is chosen to capture pitch angle scattering,
\begin{equation}
	\mathcal{C}[f] = \nu m  \frac{\norm{v_{\parallel}}}{\norm{B}} \frac{\partial}{\partial \mu} \group*{\mu \norm{v_{\parallel}} \frac{\partial f}{\partial \mu}}.
	\label{eqn:Lorenz-collision-operator}
\end{equation}
These derivatives are at fixed position and energy.
The collision frequency \(\nu\) depends only on the energy of the particle.
The velocity ratio is \(\norm{v_{\parallel}} / \norm{v} = (1 - \norm{B}/ \varrho)^{1/2}\).
The nullspace of this collision operator contains velocity-isotropic distributions, so \(\mathcal{C}[f_0 + f_1] = \mathcal{C}[f_1]\).
In this form, \eqref{eqn:Neoclassical-equation} is the linearised Lorentz-gas Fokker--Planck equation \citep[\S~13]{GoldstonRutherford1995}.

In weakly collisional plasmas, the collision frequency is small compared with the particle bounce frequency.
Consequently, fluctuations due to collisions homogenise along field lines rapidly, implying that the spatial variation in the plasma distribution along field lines in any particular magnetic well is small.
Therefore, we approximate \(f_0\) and \(f_1\) to be spatially uniform along field lines in any particular magnetic well,
\begin{gather}
	\nabla f        = \group*{\diffp{f}{\psi}}_{\alpha, \zeta, E, \mu} \nabla \psi + \group*{\diffp{f}{\alpha}}_{\psi, \zeta, E, \mu} \nabla \alpha + \group*{\diffp{f}{\zeta}}_{\psi,\alpha, E, \mu} \nabla \zeta, \\
	\norm{\nabla f} \gg \norm{(\partial f / \partial \zeta) \nabla \zeta}.
\end{gather}
Nested flux surfaces \eqref{eqn:Div-free-B1} then imply the parallel drift \(\vec{v}_{\text{Baños}}\) and the parallel spatial derivative of \(f_1\) will be negligible in the bounce-integrated drift-kinetic equation,
\begin{align}
	\overline{\mathcal{C}[f_1]} & = \nu m \frac{\partial}{\partial \mu} \mu \int \frac{\dl \zeta}{\vec{b} \vecdot \nabla \zeta} \frac{\norm{v_{\parallel}}}{\norm{B}} \diffp{f_1}{\mu} \nonumber \\
	                            & = \nu m \frac{\partial}{\partial \mu} \mu \diffp{f_1}{\mu} \overline{\norm{v_{\parallel}}^2 / \norm{B}} \nonumber                                              \\
	                            & = \frac{\partial f_0}{\partial \psi} \overline{\vec{v}_{\mathrm{D}s}\vecdot \nabla \psi}.  \label{eqn:bavg-DKE}
\end{align}
To write the last relation \eqref{eqn:bavg-DKE}, we assume there are sufficiently many passing particles so that \(f_0\) is independent of \(\alpha\),%
\footnote{The claim \(\norm{\overline{(\partial f_0 / \partial \alpha) \vec{v}_{\mathrm{D}s} \vecdot \nabla \alpha}} \ll \norm{\overline{\vec{v}_{\mathrm{D}s} \vecdot \nabla f_0}}\) requires care because \(\norm{\nabla \alpha}\) grows unbounded when the magnetic shear is non-zero. If the distribution has variation across field lines, we assume it is captured by the higher order correction \(f_1\).}
We proceed to invert the collision operator.
First, label the geodesic curvature of the field line,
\begin{equation}
	\kappa_{\mathrm{G}} = \groupbrack{\vec{b} \times (\vec{b}\vecdot \nabla \vec{b})} \vecdot \frac{\nabla \psi}{\norm{\nabla \psi}} = \frac{\vec{b} \times \nabla \norm{B}}{\norm{B}} \vecdot \frac{\nabla \psi}{\norm{\nabla \psi}}. \label{eq:kappag}
\end{equation}
The second equality is a consequence of ideal MHD force balance~\eqref{eqn:steady-state-ideal-MHD}.
Now, the primitive with respect to \(\mu\) of the bounce-integrated radial drift velocity is identified as follows:
\begin{align}
	\frac{\partial}{\partial \mu} \overline{\norm{v_{\parallel}} \beta} & = \overline{\vec{v}_{\mathrm{D}s}\vecdot \nabla \psi},                                                                                                                                                                              \\
	\diffp{\beta}{\mu}                                                  & = \frac{\vec{v}_{\mathrm{D}s}\vecdot \nabla \psi}{\norm{v_{\parallel}}} = (\norm{v}^2 \norm{v_{\parallel}}^{-1} + \norm{v_{\parallel}})  \frac{\norm{\nabla \psi} \kappa_{\mathrm{G}}}{2 \Omega_s}, \label{eqn:v_dot_grad_identity} \\
	\beta                                                               & = - (3 \norm{v}^2 \norm{v_{\parallel}} + \norm{v_{\parallel}}^3)  \frac{m \norm{\nabla \psi} \kappa_{\mathrm{G}}}{6 \Omega_s \norm{B}}. \label{eq:betadef}
\end{align}
Inverting the \(\mu\) derivative in \eqref{eqn:bavg-DKE} completes the inversion of the collision operator,
\begin{gather}
	\nu m \frac{\partial }{\partial \mu} \group*{\mu \frac{\partial f_1}{\partial \mu} \overline{\norm{v_{\parallel}}^2 / \norm{B}}} = \frac{\partial}{\partial \mu} \group*{ \frac{\partial f_0}{\partial \psi} \overline{\norm{v_{\parallel}} \beta}} \nonumber \\
	\frac{\partial f_1}{\partial \mu} = \frac{\partial f_0}{\partial \psi} \frac{\overline{\norm{v_{\parallel}} \beta}}{\nu m \mu \overline{\norm{v_{\parallel}}^2 / \norm{B}}}. \label{eqn:F1-derivative}
\end{gather}

To compute \eqref{eq:orrad}, we will use the $(E, \mu)$ parametrisation of velocity space,
\begin{align}
	\int \dl{^3\vec{v}} & = \frac{2\cpi}{m^2} \norm{B} \int_{0}^{\infty} \dl E \int_{0}^{E / \norm{B}}  \frac{\dl \mu}{\norm{v_{\parallel}}}   \label{eq:veclemu}                                               \\
	                    & = \frac{2^{1/2} \cpi}{m^{3/2}} \norm{B} \int_{0}^{\infty} \dl E \; E^{1/2}  \int_{\norm{B}}^{\infty} \frac{\dl \varrho}{\varrho^{2} (1- \norm{B}/ \varrho)^{1/2}}. \label{eq:vecerho}
\end{align}
The plasma distribution vanishes where \(\mu \geq E / \norm{B}\), so the integration region was truncated.
Using \eqref{eq:veclemu}, applying integration by parts in the $\mu$ coordinate and enforcing the boundary condition $\lim_{\mu \to 0} f_1 = 0$ at fixed energy, the radial particle flux density \eqref{eq:orrad} can be written in terms of known quantities as follows:
\begin{equation}
	\Gamma = - \int \dl{^3\vec{v}} \;  \norm{v_{\parallel}} \beta \frac{\partial f_1}{\partial \mu}. \label{eq:radpartflux2}
\end{equation}

To make optimisation efficient, the flux surface average of the radial particle flux density is of interest to minimise.
This is the average on an infinitesimal volume covering the surface,
\begin{equation}
	\mean{\Gamma} = \group*{\int \frac{\dl s}{\norm{\nabla \psi}} \Gamma} \group*{\int \frac{\dl s}{\norm{\nabla \psi}}}^{-1}. \label{eq:surfcompact}
\end{equation}
Here, \(\dl s\) is the differential surface area Jacobian.
As \eqref{eq:radpartflux2} enables computing the radial particle flux density through a quotient of bounce integrals along the magnetic field line \eqref{eqn:F1-derivative}, it is more tractable to also compute the flux surface average along the field line,
\begin{equation}
	\mean{\Gamma} = \group*{ \int_{0}^{2 \cpi} \dl \alpha \int_{\mathbb{R}} \frac{\dl \zeta}{\vec{B} \vecdot \nabla \zeta} \Gamma} \group*{\int_{0}^{2 \cpi} \dl \alpha  \int_{\mathbb{R}}  \frac{\dl \zeta}{\vec{B} \vecdot \nabla \zeta}}^{-1}. \label{eq:radpartflux}
\end{equation}

We proceed to extract a dimensionless quantity \(\Gamma_0\) for the optimisation objective.
First, we use \eqref{eq:vecerho} and \eqref{eq:radpartflux} to remove the spatial dependence in the boundary of the velocity integral,
\begin{align}
	\mean{\Gamma} & = -\frac{2 \cpi}{m^2} \group*{\int_{0}^{2 \cpi} \dl \alpha \int_{0}^{\infty} \dl E \; E \int_{\mathbb{R}}\frac{\dl \zeta}{\vec{b} \vecdot \nabla \zeta} \; \int_{\norm{B}}^{\infty} \frac{\dl \varrho}{\varrho^{2}}  \; \beta \frac{\partial f_1}{\partial \mu} }  \group*{\int_{0}^{2 \cpi} \dl \alpha \int_{\mathbb{R}} \frac{\dl \zeta}{\vec{B} \vecdot \nabla \zeta}}^{-1} \nonumber          \\
	              & = -\frac{2 \cpi}{m^2} \group*{\int_{0}^{2 \cpi} \dl \alpha \int_{0}^{\infty}  \dl E \; E \int_{\norm{B}_{\text{min}}}^{\norm{B}_{\text{max}}} \frac{\dl \varrho}{\varrho^{2}} \sum_w  \overline{\norm{v_{\parallel}} \beta} \frac{\partial f_1}{\partial \mu}} \group*{\int_{0}^{2 \cpi} \dl \alpha \int_{\mathbb{R}} \frac{\dl \zeta}{\vec{B} \vecdot \nabla \zeta}}^{-1}.  \label{eq:nospatvel}
\end{align}
Here, \(\norm{B}_{\text{min}}\) and \(\norm{B}_{\text{max}}\) are the minimum and maximum values on the flux surface.
The integral was truncated at \(\norm{B}_{\text{max}}\) as \(f_1 = 0\) for passing particles.
Now, changing coordinates in \eqref{eq:betadef},
\begin{equation}
	\beta = - \frac{(2 m E^3)^{1/2} c}{3 \varrho e \norm{B}} (1 - \norm{B} / \varrho)^{1/2} (4 \varrho / \norm{B} - 1) \norm{\nabla \psi} \kappa_{\mathrm{G}}
\end{equation}
and using the new partition for the velocity integral \eqref{eq:nospatvel}, the expression \eqref{eq:radpartflux} may be approximated using a sum over all wells in the interval \(\closeint{\zeta_1}{\zeta_2}\), as in \eqref{eq:biggamma}.

\subsection{Resolution scan for the neoclassical transport coefficient} \label{sec:neocomp}
\begin{figure}
	\centering
	\begin{subfigure}[t]{0.48\textwidth}
		\centering
		\includegraphics[width=\linewidth]{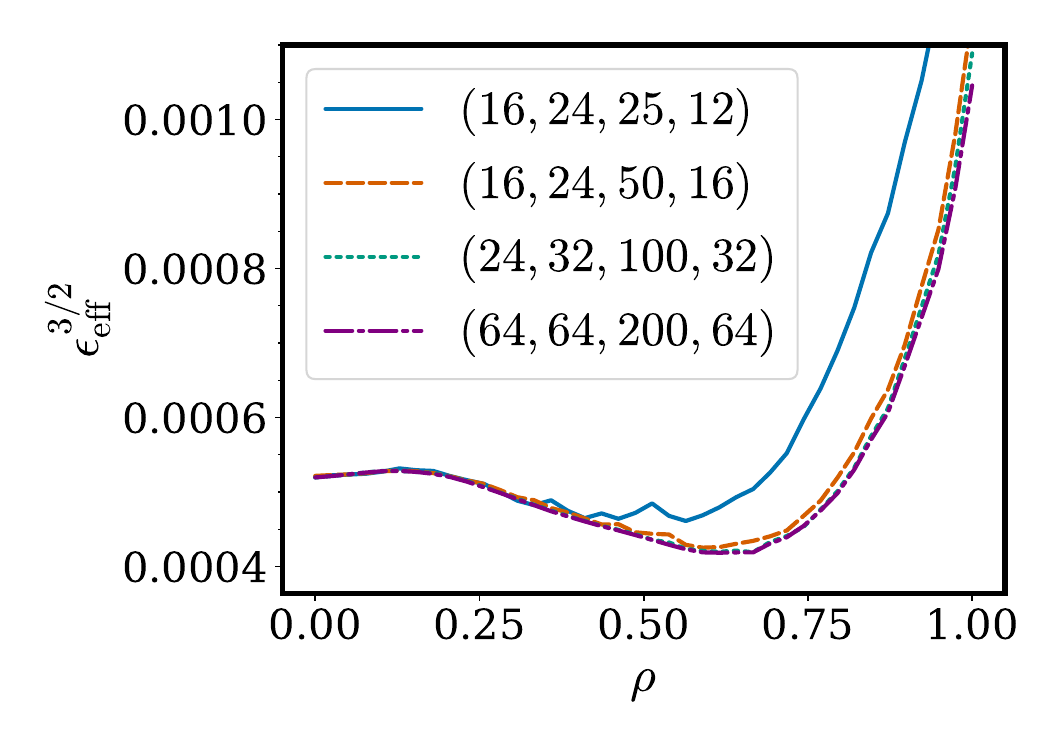}
		\caption{Five field lines are followed for \(100\) field periods. The legend shows the resolution \((X, Y, N_{\varrho}, N_q)\). \(K_{\theta} = K_{\zeta} = 33\).}
		\label{fig:DESC-res-scan}
	\end{subfigure}
	\hfill
	\begin{subfigure}[t]{0.48\textwidth}
		\centering
		\includegraphics[width=\linewidth]{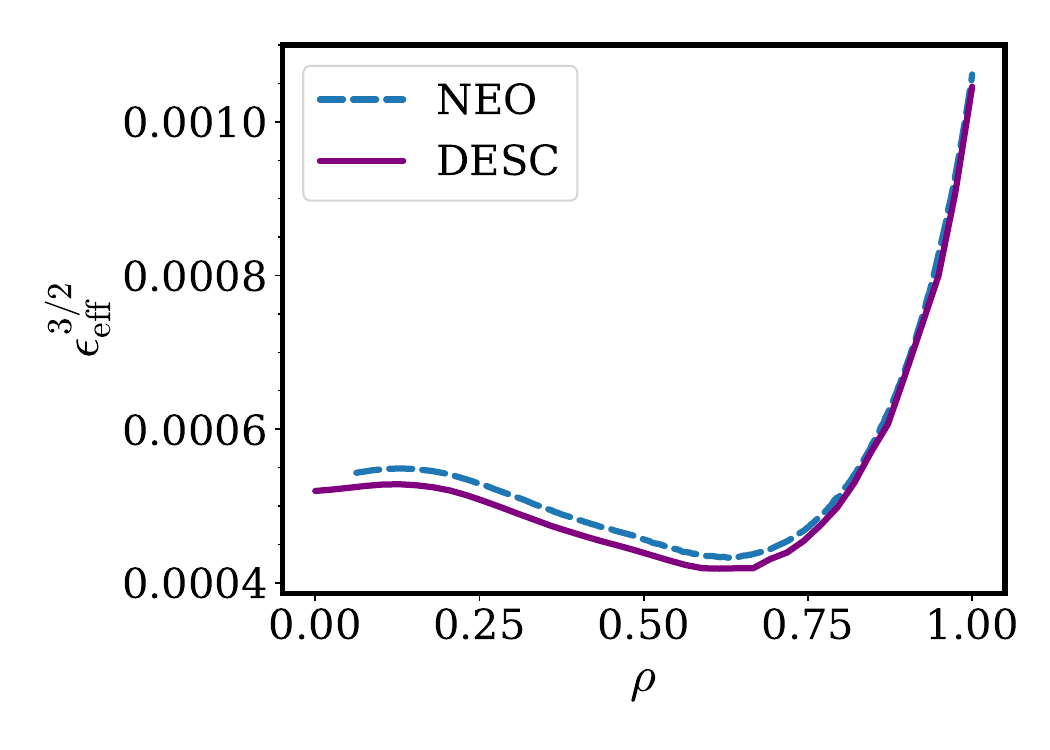}
		\caption{\texttt{NEO}-\texttt{DESC} comparison.}
		\label{fig:DESC-NEO-comparison}
	\end{subfigure}
	\caption{Resolution scan for \(\epsilon_{\text{eff}}\) on the W7-X equilibrium in the~\texttt{DESC} repository.}
\end{figure}

Figure~\ref{fig:DESC-res-scan} presents a resolution scan for $\epsilon_{\mathrm{eff}}$.
Figure~\ref{fig:DESC-NEO-comparison} compares the result to the \texttt{NEO} code~\citep{nemov1999evaluation}, which uses a finite difference technique and requires transforming to Boozer coordinates.
For the bounce integrals, \texttt{NEO} employs an explicit Runge--Kutta scheme which has algebraic convergence of order \(1 + \eta / 2\) for \eqref{eq:badint}.

We mention some performance benchmarking of our algorithm in what follows.
Computing $\epsilon_{\mathrm{eff}} \colon \mathbb{R}^{4074} \to \mathbb{R}^{10}$ and its derivative, on ten flux surfaces, following each field line for 75 field periods, with resolution \((K_{\theta}, K_{\zeta}, X, Y, N_{\varrho}, N_{q}) = (32, 32, 32, 32, 100, 32)\) was profiled to take less than one and two seconds, respectively, with a CPU \citep{intelcpu}, with peak memory consumption near one gigabyte.
These computations are at least an order of magnitude faster with a GPU.

%% file: C_drift.tex
\section{Bounce-averaged drifts in a shifted-circle model} \label{app:bavgdrift}

\begin{figure}
	\centering
	\begin{subfigure}[t]{0.48\textwidth}
		\centering
		\includegraphics[width=0.9\linewidth]{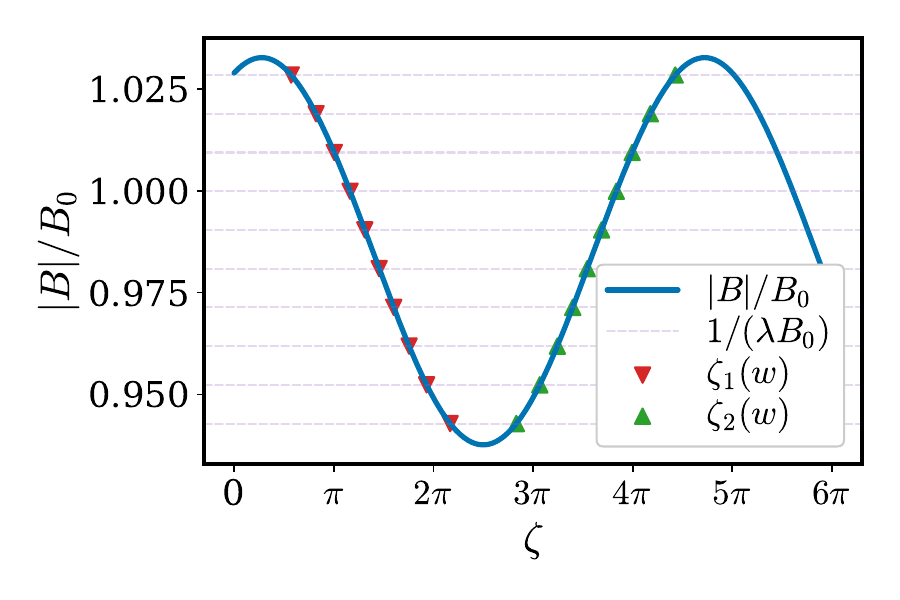}
		\caption{For a pitch marked by a horizontal line, \(\norm{v_{\parallel}} = 0\) at the points marked by triangles.}
		\label{fig:shift_well}
	\end{subfigure}
	\hfill
	\begin{subfigure}[t]{0.48\textwidth}
		\centering
		\includegraphics[width=0.9\linewidth]{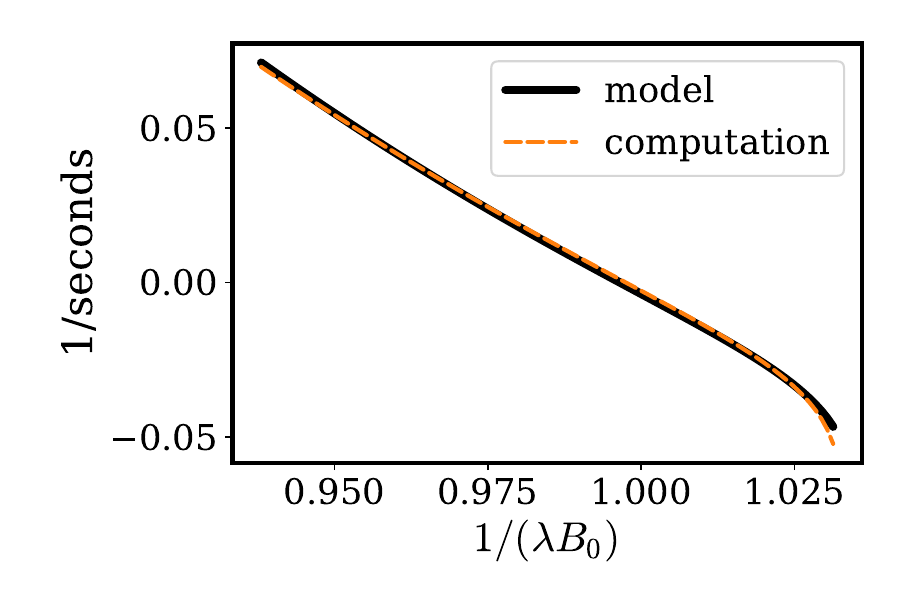}
		\caption{The bounce-averaged binormal drift in the configuration in figure \protect{\ref{fig:shift_well}} is compared.}
		\label{fig:shift_well_quad}
	\end{subfigure}
	\caption{Comparison of our shifted-circle model for the binormal drift with the result computed by our algorithm. The minor difference in panel (\protect{\subref{fig:shift_well_quad}}) is because the shifted-circle model is accurate to $\mathcal{O}(\epsilon^2)$.}
	\label{fig:driftbench}
\end{figure}

In a shifted-circle model for plasma equilibrium, we can obtain analytical expressions for bounce-averaged drifts.
We further verify our algorithm with this model in figure \ref{fig:driftbench}.

In this model, the magnetic field can be written as
\begin{equation}
	\vec{B} = \nabla \alpha \times \nabla \chi = F \nabla \phi + \diff{\chi}{\rho} \frac{\rho}{R_0} \nabla \vartheta,
\end{equation}
where $\alpha = \phi - \iota^{-1} \vartheta$, \(\chi\) is the poloidal flux, $F$ is the enclosed poloidal current, \(R_0\) is the average major radius and \(\rho\) is a radial coordinate.
To lowest order, the Grad--Shafranov equation has the constant solution $F = F_0$.
To the next order, the pressure gradient is
\begin{equation}
	\dl p / \dl \rho = - F_0 R^{-2} \diffs{F}{\rho}.
\end{equation}
To first order, the poloidal magnetic field can be ignored so that the field satisfies $\norm{B} = B_0 (1 - \epsilon \cos \vartheta)$ and \(\vec{b}\vecdot \nabla \vartheta = G_0 (1-\epsilon\cos  \vartheta)\), where $\epsilon \ll 1$ is the reciprocal of the aspect ratio. Here, \(B_0\) and \(G_0\) are constants.
In this model, the global shear \(\hat{s}\), normalised pressure gradient and integrated local shear are
\begin{align}
	\hat{s}               & = - \rho \iota^{-1} \diffs{\iota}{\rho},                                                                                                                                                     \\
	\alpha_{\mathrm{MHD}} & = -2^{-1} \iota^{-2} \diffs{p}{\rho},                                                                                                                                                        \\
	\texttt{gds21}        & = \diffs{\chi}{\rho} \diffs{\iota^{-1}}{\rho} \nabla \chi \vecdot \nabla \alpha = -\hat{s} (\hat{s}\vartheta-  \norm{B}^{-4} \alpha_{\mathrm{MHD}} \sin \vartheta ) + \mathcal{O}(\epsilon).
	\label{eq:gds21}
\end{align}
The binormal, geometric part of the \(\nabla \norm{B}\) drift is
\begin{equation}
	(\nabla \norm{B})_{\text{drift}} = \norm{B}^{-3} (\vec{B}\times \nabla \norm{B}) \vecdot \nabla \alpha = f_{2} (-\hat{{s}}+ \cos \vartheta- \hat{s}^{-1} \texttt{gds21} \sin \vartheta ).
	\label{eqn:gbdrift}
\end{equation}
The binormal, geometric part of the curvature drift is
\begin{align}
	\texttt{cvdrift} & = \norm{B}^{-3} [\vec{B}\times \nabla (p + \norm{B}^2 / 2) ] \vecdot \nabla \alpha                                                                                                                            \nonumber  \\
	                 & = (\nabla \norm{B})_{\text{drift}} +f_{3} \norm{B}^{-3} \diffs{p}{\rho}                                                                                                                                        \nonumber \\
	                 & =f_{2} (-\hat{s}+ \cos \vartheta+\hat{s}\vartheta\sin \vartheta- B_0^{-4} \alpha_{\mathrm{MHD}} \sin^2 \vartheta  )+f_{3} B_0^{-2} \alpha_{\mathrm{MHD}} + \mathcal{O}(\epsilon).
	\label{eqn:cvdrift}
\end{align}
The scalars $f_2$ and $f_3$ contain some constants.
The bounce-averaged drift is
\begin{equation}
	\mean{v_{\mathrm{{D}}}} = \group*{ \int_{\vartheta_1}^{\vartheta_2}\frac{\dl\vartheta}{\vec{b}\vecdot\nabla\vartheta}\norm{v_{\parallel}}^{-1} }^{-1} \int_{\vartheta_1}^{\vartheta_2}\frac{\dl\vartheta}{\vec{b}\vecdot\nabla\vartheta} \groupbrack*{\norm{v_{\parallel}}\texttt{{cvdrift}}+\frac{\norm{v_{\perp}}^{2}}{2 \norm{v_{\parallel}}}  (\nabla \norm{B})_{\text{drift}}}.
	\label{eq:bavg-drift}
\end{equation}

As used by \cite{Connor_1983} and shown by \cite{10.1063/1.4926818}, in the limit of a large aspect ratio shifted-circle model, the parallel speed of a particle with a fixed energy is \(
\norm{v_{\parallel}} = (2E/m)^{1/2} (2\epsilon \lambda B_0)^{1/2} ({k^{2}-\sin^2 (\vartheta/2)})^{1/2}\),
where
\begin{equation}
	k^{2}=2^{-1} [(1-\lambda B_{0})(\epsilon\lambda B_{0})^{-1} +1]
\end{equation}
parametrises the pitch angle.
Using these simplifications and $\norm{v_{\perp}}^2/2 = E/m - \norm{v_{\parallel}}^2/2$,
\begin{align}
	\mean{v_{D}} & = \left(\int_{-2\arcsin k}^{2\arcsin k}\frac{{\dl\vartheta}}{\vec{b}\vecdot\nabla\vartheta} (2 \epsilon \lambda B_0)^{-1/2} (k^{2}-\sin^2(\vartheta/2))^{-1/2} \right)^{-1} \nonumber     \\
	             & \quad \int_{-2\arcsin k}^{2\arcsin k}\frac{{\dl\vartheta}}{\vec{b}\vecdot\nabla\vartheta}\Big[(2 \epsilon \lambda B_0)^{1/2} (k^{2}-\sin^2(\vartheta/2))^{1/2} \texttt{cvdrift} \nonumber \\
	             & \qquad -  2^{-1/2} (\epsilon \lambda B_0)^{1/2} (k^{2}-\sin^2(\vartheta/2))^{1/2} (\nabla \norm{B})_{\text{drift}} \nonumber                                                              \\
	             & \qquad +  2^{-3/2} (\epsilon \lambda B_0)^{-1/2} (k^{2}-\sin^2(\vartheta/2))^{-1/2} (\nabla \norm{B})_{\text{drift}} \Big].
	\label{eqn:bounce-average-drift-analytical}
\end{align}

The following identities simplify~\eqref{eqn:bounce-average-drift-analytical}.
The incomplete elliptic integrals are converted to complete elliptic integrals using the reciprocal-modulus transformation in \eqref{eqn:I0} and \eqref{eqn:I1} \citep{DLMF}.
Here, $K$ and $E$ are complete elliptic integrals of the first
and second kind, respectively,
\begin{align}
	\mathsf{I_{0}} & = \int_{-2\arcsin k}^{2\arcsin k} \dl\vartheta \; ({k^{2}-\sin^2 (\vartheta/2)})^{-1/2} = 4 K(k)
	\label{eqn:I0},                                                                                                                                                                                \\
	\mathsf{I_{1}} & =\int_{-2\arcsin k}^{2\arcsin k}\dl\vartheta \; ({k^{2}-\sin^2 (\vartheta/2)})^{1/2} = 4 \left[E(k) + (k^2-1) K(k) \right]
	\label{eqn:I1},                                                                                                                                                                                \\
	\mathsf{I_{2}} & = \int_{-2\arcsin k}^{2\arcsin k} \dl\vartheta \; (k^{2}-\sin^2 (\vartheta/2))^{-1/2} \vartheta \sin \vartheta = 16 \left[E(k) + (k^2-1) K(k) \right],                        \\
	\mathsf{I_{3}} & =\int_{-2\arcsin k}^{2\arcsin k} \dl \vartheta \; (k^{2}-\sin^2 (\vartheta/2))^{1/2} \vartheta \sin \vartheta =\frac{16}{9}\left[(4 k^2-2) E + (3 k^4 - 5 k^2 + 2) K \right], \\
	\mathsf{I_{4}} & = \int_{-2\arcsin k}^{2\arcsin k} \dl\vartheta \; ({k^{2}-\sin^2 (\vartheta/2)})^{-1/2} \sin^2 \vartheta = \frac{16}{3} [(2k^2 -1) E +  (1 - k^2) K ],                        \\
	\mathsf{I_{5}} & = \int_{-2\arcsin k}^{2\arcsin k} \dl\vartheta \; (k^{2}-\sin^2 (\vartheta/2))^{1/2} \sin^2 \vartheta  \nonumber                                                              \\
	               & \quad = \frac{16}{15} \left[  2(1 - k^2 + k^4) (E + (k^2-1) K)  - (1 - 3 k^2 + 2 k^4) k^2 K \right],                                                                          \\
	\mathsf{I_{6}} & = \int_{-2\arcsin k}^{2\arcsin k} \dl \vartheta \; (k^{2}-\sin^2 (\vartheta/2))^{-1/2} \cos \vartheta = 8 E - 4 K,                                                            \\
	\mathsf{I_{7}} & = \int_{-2\arcsin k}^{2\arcsin k} \dl \vartheta \; (k^{2}-\sin^2 (\vartheta/2))^{1/2} \cos \vartheta  =\frac{4}{3} \left[(2k^2-1) E - (k^2-1) K \right].
\end{align}
Using these formulae with \(\epsilon_{\lambda} = 2 \epsilon \lambda B_0\), to lowest order, the bounce-averaged drift is
\begin{align}
	\mean{v_{D}} & = \frac{1}{\mathsf{I_{0}}} \groupbrack*{f_3 \frac{\alpha_{\mathrm{MHD}}}{B_0^2} \epsilon_{\lambda} \mathsf{I}_1 - \frac{f_2}{2} \group*{\hat{s} \left( \mathsf{I}_0 + \epsilon_{\lambda} \mathsf{I}_1 - \mathsf{I}_2 - \epsilon_{\lambda} \mathsf{I}_3 \right) + \frac{ \alpha_{\mathrm{MHD}}}{B_0^4} (\mathsf{I_{4}} + \epsilon_{\lambda} \mathsf{I_{5}}) - (\mathsf{I_{6}} + \epsilon_{\lambda} \mathsf{I_{7}})}}.
\end{align}

%% file: D_pest_transform.tex
\section{Issues with changing the spectral basis to straight field line coordinates} \label{specbasistopest}
\begin{figure}
	\centering
	\begin{subfigure}{0.975\textwidth}
		\centering
		\includegraphics[width=0.9\linewidth]{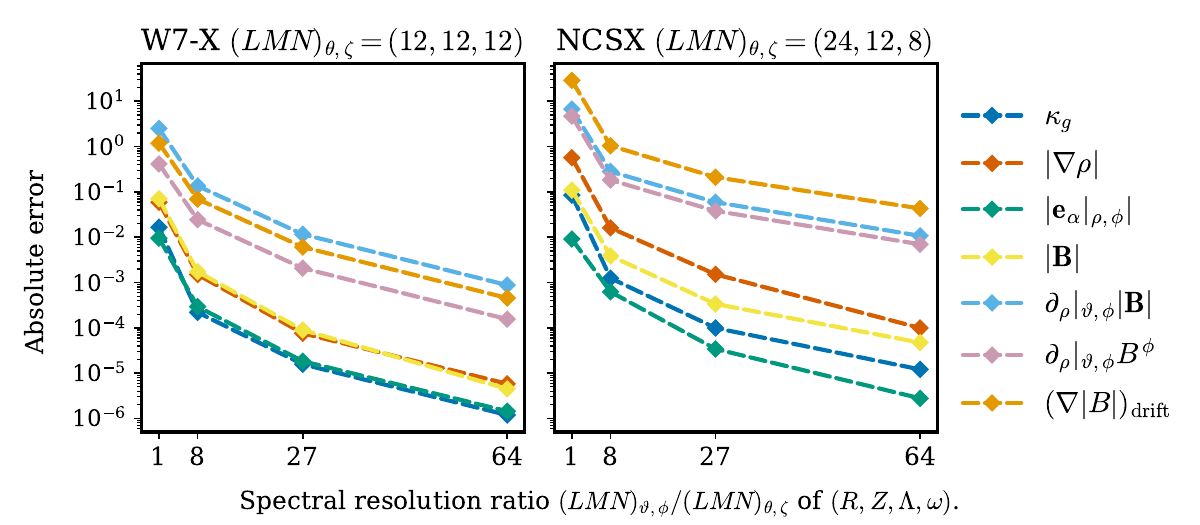}
		\caption{Equation \protect{\eqref{eq:alphamap}} was solved to error \(\leq 10^{-10}\) on the tensor-product of the optimal concentric sampling grid \protect{\citep{RAMOSLOPEZ2016247}} in \((\rho, \vartheta) \in \closeint{0}{1} \times \clopenint{0}{2 \cpi}\) and a uniform grid in \(\phi \in \clopenint{0}{2 \cpi / N_{\text{FP}}}\). \((R, Z, \Lambda, \omega)\) were interpolated to a Fourier--Zernike series in \((\rho, \vartheta, \phi)\) with maximum mode numbers \((L, M, N)_{\vartheta, \phi}\) on this grid.
			The interpolation used a \(1.5 \times\) over-sampled, in both \(\rho\) and \(\vartheta\), weighted least-squares fit to improve conditioning for the Zernike series, followed by an FFT in \(\phi\).
			(The optimal grid for interpolation to a Zernike series does not coincide with the optimal grid for quadrature to project onto the Zernike basis because the Zernike basis is not a tensor-product basis. Interpolation with a weighted least-squares fit was chosen because the interpolation grid is sparser than the quadrature grid.)
			Each quantity was then computed on a uniform grid in \((\rho, \vartheta, \phi)\). Quadrature required to compute a quantity in the plot was done on an over-sampled grid to account for nonlinearity in the computation from \((R, Z, \Lambda, \omega)\).
			Zernike polynomials were evaluated with stable Jacobi polynomial recurrence relations using the algorithm in \protect{\cite{ELMACIOGLU2025129534}}.}
		\label{fig:pestconvert}
	\end{subfigure}
	\begin{subfigure}{0.975\textwidth}
		\centering
		\includegraphics[width=0.9\linewidth]{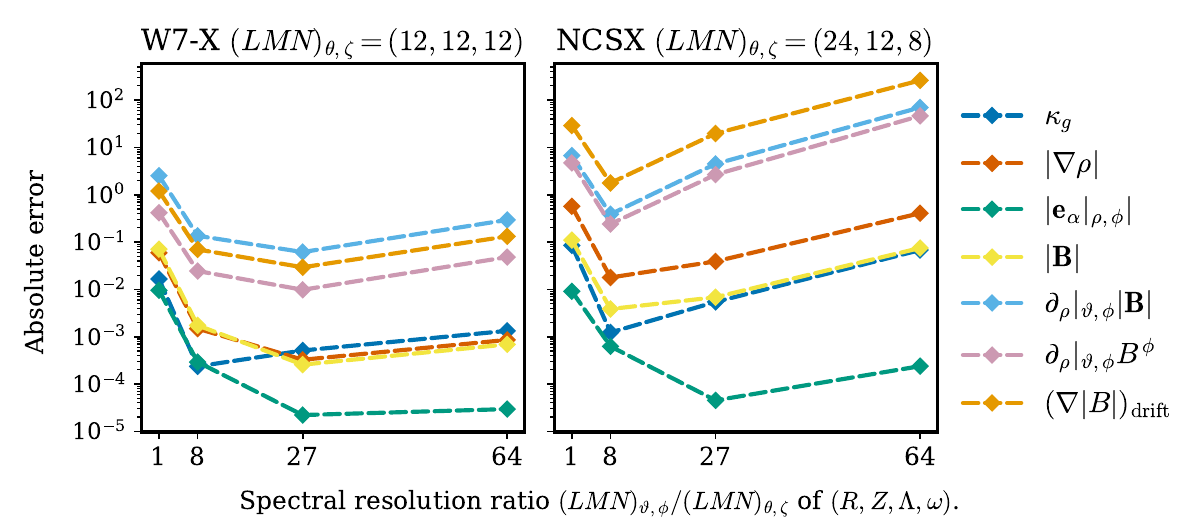}
		\caption{This is the same demonstration as figure \protect{\ref{fig:pestconvert}} except equation \protect{\eqref{eq:alphamap}} is solved to error \(10^{-7}\).}
		\label{fig:pestconvertbad}
	\end{subfigure}
	\caption{Error induced by changing the Fourier--Zernike basis for \((R, Z, \Lambda, \omega)\) from flux coordinates \((\theta, \zeta)\) to the straight field line coordinates \((\vartheta, \phi)\). Fitting at the resolution that obtains the error of \(10^{-4}\) Tesla in \(\norm{B}\) on the NCSX stellarator in panel (\protect{\subref{fig:pestconvert}}) took 10 minutes with a CPU \citep{intelcpu}.}
	\label{fig:lastfig}
\end{figure}

In figure \ref{fig:lastfig}, we show that parametrising our basis for \((R, Z, \Lambda, \omega)\) in straight field line coordinates \((\vartheta, \phi) = (\theta + \Lambda, \zeta + \omega)\) \citep{PEST2012} is inefficient and ill-conditioned.
Therefore, we use the approach discussed in the main text instead.
